# A Stochastic Differential Equation Framework for Guiding Online User Activities in Closed Loop

**Yichen Wang**
Georgia Tech

**Evangelos Theodorou**
Georgia Tech

**Apurv Verma**
Georgia Tech

**Le Song**
Georgia Tech & Ant Financial

## Abstract

Recently, there is a surge of interest in using point processes to model continuous-time user activities. This framework has resulted in novel models and improved performance in diverse applications. However, most previous works focus on the "open loop" setting where learned models are used for predictive tasks. Typically, we are interested in the "closed loop" setting where a policy needs to be learned to incorporate user feedbacks and guide user activities to desirable states. Although point processes have good predictive performance, it is not clear how to use them for the challenging closed loop activity guiding task. In this paper, we propose a framework to reformulate point processes into stochastic differential equations, which allows us to extend methods from stochastic optimal control to address the activity guiding problem. We also design an efficient algorithm, and show that our method guides user activities to desired states more effectively than state-of-arts.

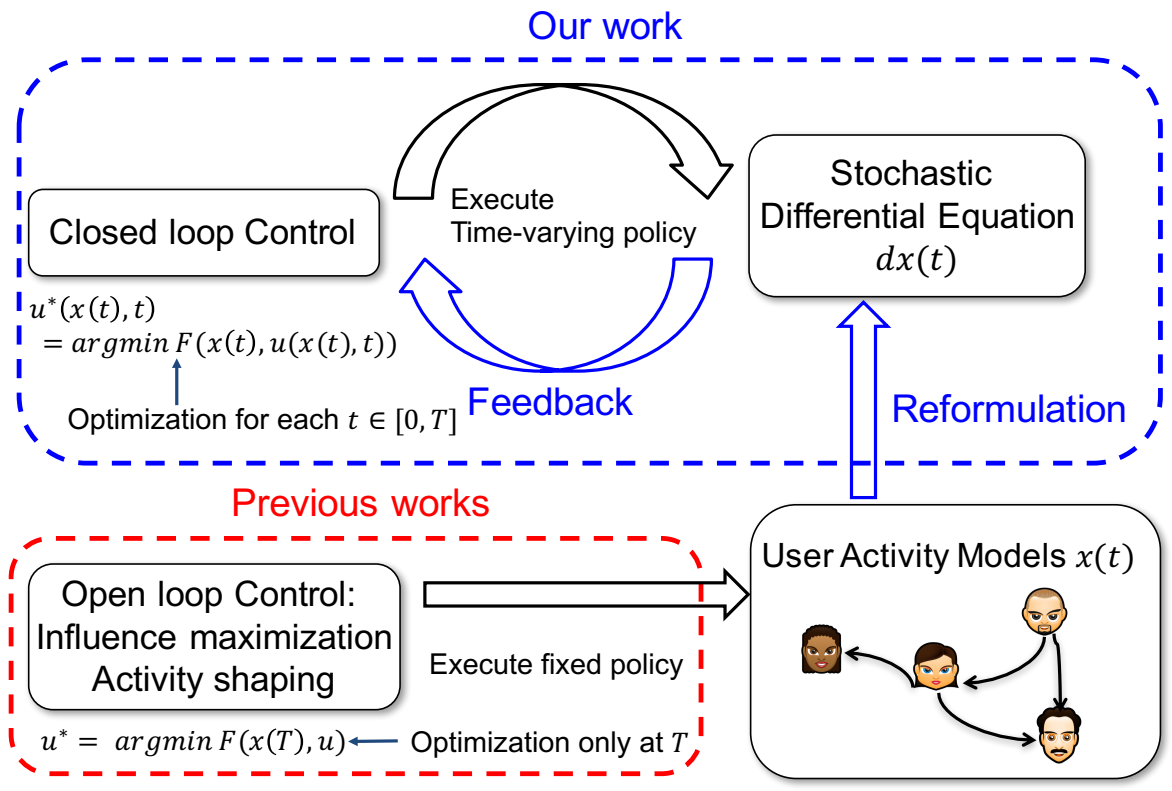

Figure 1: Comparison between our work and previous works in guiding user activities. Given a user activity model that is learned from observed behaviors, our "closed loop" guiding framework aims to find an optimal policy $u(x(t), t) : \Re \times \Re \to \Re$ that maps a user's current state $x(t)$, *e.g.*, opinion, to an action. On the contrary, previous works are "open loop": they only optimize the objective function at terminal time and compute a fixed scalar policy $u \in \Re$.

## 1 Introduction

Online social and information platforms have brought to the public a new style of social lives: people use these platforms to receive information, create content, and share opinions. The large-scale temporal event data generated by online users have created new research avenues and scientific questions at the intersection of social sciences and machine learning. These questions are directly related to the development of new models as well as learning and inference algorithms to understand and predict user activities from these data (Kempe et al., 2003; Zhao et al., 2015; Grover & Leskovec, 2016; Lian et al., 2015; Nowzari et al., 2016; Pan et al., 2016; Tan et al., 2016; He & Liu, 2017).

Recently, point processes have been widely applied to model user activities. Instead of discretizing time into intervals, these models treat timestamps as random variables, and propose a mechanism to link model parameters with the observed timestamps. Such fine-grained modeling of temporal information has led to improved predictive performance in diverse applications, such as information diffusion (Du et al., 2013; Farajtabar et al., 2014; He et al., 2015; Wang et al., 2016a,b), recommender systems (Du et al., 2015; Dai et al., 2016; Wang et al., 2016a), and evolutionary networks (Farajtabar et al., 2015). However, most works deal with the "open loop" setting where models are used mainly for prediction, but user feedbacks are not incorporated into the model. Typically, we are interested in the "closed loop" setting where we want to design a policy to guide user activities and incorporate feedbacks timely. It is not clear how the current models can be used for such "closed loop" task.

For instance, a decision maker seeks to determine the best intervention policy such that the sentiment in user generated contents can be guided towards a positive state. This is significantly more challenging than traditional influence maximization or activity shaping tasks (Chen et al., 2009, 2010a,b, 2012; Farajtabar et al., 2014), where the policy is determined before the process unfolds, and they do not take into account the instantiation of the process. A framework for doing this is critically important for understanding the vulnerabilities of information networks, designing policies to suppress the spread of undesired contents, and exploring new ways of performing content recommendation. For instance, a network moderator may want to effectively contain the spread of rumors and misinformation, and an online platform may want to promote the level of long-term user activities rather than the short-term click-through rate.

In this paper, we provide a novel view of point process models, and reformulate them into stochastic differential equations (SDEs). This reformulation allows us to significantly generalize existing point process models to SDEs, and plays a critical role in connecting the task of guiding user activities to stochastic optimal control theory, which are often used in robotics. Hence, we can bring in and extend lots of tools from stochastic optimal control literature to address the "closed loop" sequential decision making problem. Figure 1 illustrates our framework.

Interestingly, these problems also introduce new technical challenges. Previous works in stochastic optimal control study SDEs driven by Wiener processes and/or Poisson processes (Boel & Varaiya, 1977; Pham, 1998; Oksendal & Sulem, 2005; Hanson, 2007). Online user activity modeling requires us to consider more advanced processes, such as (i) Hawkes processes for long term memory and mutual exciting phenomena in social interactions, (ii) survival processes (Aalen et al., 2008) for self-terminating behavior in influence propagation and link creation, and (iii) node birth processes for evolving networks (Farajtabar et al., 2015; Thalmeier et al., 2016). Thus, many technical results from control theory need to be extended for the activity guiding problem. In summary, we make the following contributions:

- We provide a general way to reformulate point process based user activity models into stochastic differential equations (SDEs), which allows us to bring in and extend tools from stochastic control literature to address the "closed loop" activity guiding problem.
- We extend technical results in stochastic optimal control theory and derive generalized Ito's lemma and HJB equation for SDEs driven by more general point processes.
- We propose an algorithm that efficiently guides the dynamics of user activities towards a target. The algorithm can also deal with the challenging cases of time-varying networks with node births.

Finally, with synthetic and real networks, we showed our framework is robust, able to steer user activities to desired states with faster convergence rate and less variance than the state-of-art methods.

**Further related work**. Most previous works in influence maximization (Kempe et al., 2003; Chen et al., 2010b, 2012) focus on selecting sources to maximize the spread of information in infinite time, which is an "open loop" setting. There is typically no quantitative prescription on how much incentive should be provided to each user. Moreover, in most cases, a finite time window must be considered. For example, a politician would like to maximize his support by a million people in one week instead of fifty years.

Another relevant area is optimal control for epidemic processes (Epanchin-Niell & Wilen, 2012; Nowzari et al., 2016; Ogura & Preciado, 2016; Pastor-Satorras et al., 2015). However, the epidemic processes are modeled by deterministic differential equations. Therefore, these works neither consider the influence of abrupt event nor the stochastic nature of user behaviors in social networks. Hawkes process models (Farajtabar et al., 2014; De et al., 2015) overcome this limitation by treating the user-generated event time as a random variable and model users' stochastic behaviors. They address the problem of designing the base intensity of Hawkes process to achieve a desired steady behavior. However, these methods are open loop and do not consider user feedbacks, and the variance of the dynamics can still be very large. Recently, Thalmeier et al. (2016) studied the problem of controlling network growth in the macroscopic level, while we focus on guiding users' microscopic temporal behaviors based on point processes and SDEs.

## 2 Background and preliminaries

**Point processes**. A temporal point process (Aalen et al., 2008) is a random process whose realization is a list of events $\{t_i\}$, localized in time. Many types of data in social networks can be represented as point processes (Du et al., 2015; Farajtabar et al., 2014; Lian et al., 2015; Tan et al., 2016; Wang et al., 2016a,b, 2017a,b,c). The point process can be equivalently represented as a counting process, $N(t)$, which records number of events before time $t$. An important way to characterize point processes is via the conditional intensity function $\lambda(t)$, which is a stochastic model for the time of the next event given all previous events. Let $\mathcal{H}(t) = \{t_i | t_i < t\}$ be the history of events happened up to $t$. Formally, $\lambda(t)$ is the conditional probability of observing an event in a small window $[t, t+dt)$ given the history $\mathcal{H}(t)$: $\lambda(t)\mathrm{d}t := \mathbb{P}\{\text{event in } [t, t+\mathrm{d}t) | \mathcal{H}(t)\} = \mathbb{E}[\mathrm{d}N(t)|\mathcal{H}(t)]$, where only one event can happen in a window of $\mathrm{d}t$, *i.e.*, $\mathrm{d}N(t) \in \{0, 1\}$. The functional form of $\lambda(t)$ is often designed to capture the phenomena of interests. We will provide more examples later in Section 3.

**Stochastic differential equations (SDEs).** A jump diffusion SDE is a differential equation in which one or more terms is a stochastic process: $\mathrm{d}x(t) = f(x)\mathrm{d}t + g(x)\mathrm{d}w(t) + h(x)\mathrm{d}N(t)$, where $\mathrm{d}x(t) := x(t+\mathrm{d}t) - x(t)$ is the differential of $x(t)$. This SDE contains a drift, a diffusion and a jump term: the drift term $f(x)\mathrm{d}t$ models the intrinsic evolution of $x(t)$; the diffusion Wiener process $w(t) \sim \mathcal{N}(0, t)$ captures the Gaussian noise; the jump point process $N(t)$ captures the influence of abrupt events.

## 3 Stochastic differential equations for user activity models

In this section, we establish a framework to reformulate many point process models into SDEs. This reformulation framework plays a critical role in connecting the task of guiding user activities to stochastic optimal control theory often used in robotics and designing closed loop policies.

### 3.1 User activity models

We first introduce the generic point process models for user activities, and present three examples.

**Definition 1** (User Activity Model)**.** *For a network with $U$ users, the point process $N_i(t)$ models the generation of event times from user i, and its intensity f is defined as*

$$\lambda_i(t) = \eta_i(t) + \sum\nolimits_{j=1}^{U} \beta_{ij} \sum\nolimits_{t_j \in \mathcal{H}_j(t)} \kappa_{\omega_1}(t - t_j), \quad (1)$$

*where $\eta_i(t)$ is the base intensity, $\beta_{ij} \geqslant 0$ models the strength of influence from user $j$ to $i$, $\mathcal{H}_j(t)$ is the history of events for user $j$, and $\kappa_{\omega_1}(t) = \exp(-\omega_1 t)$ is the triggering kernel capturing the influence of past event. We also assume the additional event feature/content $x_i(t)$ follows the model*

$$x_i(t) = b_i(t) + \sum\nolimits_{j=1}^{U} \alpha_{ij} \sum\nolimits_{t_j \in \mathcal{H}_j(t)} \kappa_{\omega_2}(t - t_j)\, h(x_j(t_j)) \quad (2)$$

*where $b_i(t)$ is the base content, $\alpha_{ij}$ is the influence from user $j$ to $i$, the function $h(\cdot)$ captures the influence of the activity content and typical forms include $h = 1$ and $h(x) = x$.*

This model generalizes point processes since it not only models the on-off temporal behavior, but also the activity content $x_i(t)$, *e.g.*, opinion or a vector of interested topics. It captures both exogenous and endogenous properties of networks. The exogenous term is the base intensity/activity content, and the endogenous term captures the fact that one's intensity/activity content is influenced by neighbors. $\alpha_{ij}, \beta_{ij}$ measure the strength of such influence, the kernel $\kappa_\omega(t)$ captures the decay of influence of past events/content over time, and the summation captures the influence of neighbors' events. Next we present three examples.

- **Continuous-time information propagation** (Du et al., 2013). The information propagation in social networks begins with a set of source nodes, and the contagion is transmitted from the sources along their out-going edges to their direct neighbors. We set $N_{ij}(t)$ to be the *survival process* capturing the infection on the edge $i \to j$, and $N_{ij}(t) = 1$ means node $j$ is infected by node $i$ at time $t$. Since there is only one event for an instantiation of this process, we set the infection intensity as follows:
$$\lambda_{ij}(t) = \eta_{ij}(1 - N_{ij}(t)) \quad (3)$$
where $\eta_{ij}$ is intensity that $i$ infects $j$, and $(1 - N_{ji}(t))$ ensures the infection happens only once.
- **Hawkes processes** (Hawkes, 1971). This model captures the mutual excitation phenomena between events, and has been successfully applied to analyze user behaviors in social networks (Du et al., 2015; Farajtabar et al., 2014, 2015; Wang et al., 2016a,b, 2017b). The process $N_i(t)$ counts the number of events generated by user $i$ up to time $t$ and its intensity $\lambda_i(t)$ models mutual excitation between the events from a collection of $U$ users as follows:
$$\lambda_i(t) = \eta_i + \sum\nolimits_{j=1}^{U} \beta_{ij} \sum\nolimits_{t_j \in \mathcal{H}_j(t)} \kappa_{\omega_1}(t - t_j)$$
Here, the occurrence of each historical event from one's neighbors increases the intensity by a certain amount determined by $\kappa_{\omega_1}(t)$ and $\alpha_{ij}$.
- **Opinion dynamics** (De et al., 2015; He et al., 2015). This model considers both the timing and content of each event. It assigns each user $i$ a Hawkes intensity $\lambda_i(t)$ to generate events and an opinion process $x_i(t)$, where $x_i(t) = 0$ corresponds to neutral opinion and large positive/negative values correspond to extreme opinions. The opinion of user $i$ is modeled as a temporally discounted average of neighbors' opinion:
$$x_i(t) = b_i + \sum\nolimits_{j} \alpha_{ij} \sum\nolimits_{t_j \in \mathcal{H}_j(t)} \kappa_{\omega_2}(t - t_j) x_j(t_j)$$
This model has superior performance in predicting opinions. However, it is not clear whether it can be used to design feedback policies to guide the dynamics precisely to some target states.

### 3.2 Equivalent SDE reformulation

We are now ready to show the novel SDE reformulation of the user activity model in Definition 1.

**Theorem 2** (Transformation Framework)**.** *The equivalent SDE form of the user activity model is:*

$$\mathrm{d}\lambda_i(t) = \mathrm{d}\eta_i + \omega_1\big(\eta_i - \lambda_i\big)\mathrm{d}t + \sum\nolimits_{j} \beta_{ij} \mathrm{d}N_j(t) \quad (4)$$
$$\mathrm{d}x_i(t) = \mathrm{d}b_i + \omega_2\big(b_i - x_i(t)\big)\mathrm{d}t + \sum\nolimits_{j} \alpha_{ij} h(x_j) \mathrm{d}N_j(t)$$

**Proof sketch**. We first define a convolution operator and reformulate the user activity model in (1) and (2), next we apply a differential operator d to the reformulated equations and derive the differential form of $\lambda_i(t)$ and $x_i(t)$, which lead to two SDEs. Appendix A contains details.

These SDEs describe how the intensity $\lambda_i(t)$ and content $x_i(t)$ change on $[t, t + \mathrm{d}t)$; each consists of three terms:

- **Baseline change**. The differential $\mathrm{d}b_i(t)$ captures the infinitesimal change of base activity content.
- **Drift**. The change rate of the activity content, $\mathrm{d}x_i(t)/\mathrm{d}t$, is proportional to $-x_i(t)$, which means the activity content $x_i(t)$ tends to stabilize over time. In fact, if ignoring the jump term and $\mathrm{d}b_i(t)$, the expected activity content $\mathbb{E}[x_i(t)]$ will converge to $b_i(t)$ as $t$ increases. This can be proved by setting $\mathbb{E}[\mathrm{d}x_i(t)] = 0$. Similarly, equation (4) shows a user's intensity tends to stabilize over time.
- **Jump**. This term captures the influence of each event in the network and is a weighted summation of the neighbors' influence. The coefficient $\alpha_{ij}$ ensures that only neighbors' effect will be considered, and $\mathrm{d}N_j(t) \in$ $\{0, 1\}$ models whether user $j$ generates an event. Similarly, equation (4) shows that user $i$'s intensity increases by $\beta_{ij}$ if his neighbor $j$ generates an event.

Next, we present three applications of Theorem 2.

- **SDE for continuous-time information propagation**. The intensity in (3) is a simplified version of (1) without the historical influence term. Its SDE version is:

$$\mathrm{d}\lambda_{ij}(t) = -\eta_{ij}\mathrm{d}N_{ij}(t)$$

This SDE keeps the key property of survival process: before an infection happens, the intensity is constant, *i.e.*, $\lambda_{ij}(t) = \eta_{ij}$; after the infection happens, the intensity is 0, *i.e.*, $\lambda_{ij}(t + \mathrm{d}t) = \lambda_{ij}(t) + \mathrm{d}\lambda_{ij}(t) = 0$.

- **SDE for Hawkes process**. We set $\eta_i(t) = \eta_i$ and obtain:

$$\mathrm{d}\lambda_i(t) = \omega_1\big(\eta_i - \lambda_i(t)\big)\mathrm{d}t + \sum\nolimits_j \alpha_{ij}\mathrm{d}N_j(t)$$

This SDE shows that the user's intensity tends to stabilize and its change is influenced by his neighbors' activities.

- **SDE for opinion dynamics.** We set $b_i(t) = b_i$, $h_j(x_j) = x_j$, and further generalize this model by adding a Wiener process term:

$$\mathrm{d}x_i = \omega_2\big(b_i - x_i\big)\mathrm{d}t + \beta\mathrm{d}w_i + \sum\nolimits_j \alpha_{ij}x_j\mathrm{d}N_j$$

where the Wiener process $w_i(t)$ captures the Gaussian noise, such as fluctuations in the dynamics due to unobserved factors and activities outside the social platform. The jump term models the fact that the change of opinion is influenced by his neighbors' opinion.

### 3.3 Benefit of the SDE modeling framework

Our SDE formulation opens a new gate to extend tools from optimal control theory. Hence we can solve many important problems in social sciences, such as the least square activity guiding and activity maximization problem. Without this view, it is not easy to design algorithms with closed loop policies. Besides transforming an existing model to an SDE, we can also directly design an SDE to model many other factors. For example, one can model the Gaussian noise by adding a Wiener process to an SDE. Next, we show how to optimally control the SDE to guide user activities.

## 4 A convex activity guiding framework

In this section, we define the activity guiding problem. Let $\boldsymbol{x}(t) = (x_1(t), \cdots, x_U(t))^\top$ be the vector of each user's activity content, then we study the vector version of the SDE in (4) and reformulate this SDE with an extra control policy $\boldsymbol{u}(\boldsymbol{x}(t), t) : \Re^U \times \Re \to \Re^U$ as follows.

$$\mathrm{d}\boldsymbol{x} = (\boldsymbol{f}(\boldsymbol{x}) + \boldsymbol{u})\mathrm{d}t + \boldsymbol{g}(\boldsymbol{x})\mathrm{d}\boldsymbol{w}(t) + \boldsymbol{h}(\boldsymbol{x})\mathrm{d}\boldsymbol{N}(t) \quad (5)$$

This policy $\boldsymbol{u}$ maps the activity content $\boldsymbol{x}$ to an action for all users. Next we present two examples.

- **Guiding Hawkes process.** To steer user activities to a desired level, we provide incentives to users and add a control policy $u_i(\lambda_i(t), t)$ to the SDE formulation of the Hawkes intensity function as follows:

$$\mathrm{d}\lambda_i(t) = (\eta_i + u_i(\lambda_i(t), t) - \lambda_i(t))\mathrm{d}t + \sum\nolimits_j \beta_{ij}\mathrm{d}N_j(t),$$

where $u_i(\lambda_i(t), t)$ captures the amount of additional influence to change the baseline intensity $\eta_i$.

- **Guiding opinion dynamics.** We can guide the opinion SDE with a policy $u_i(x_i(t), t)$ as follows:

$$\mathrm{d}x_i = (b_i + u_i - x_i)\mathrm{d}t + \theta\mathrm{d}w_i + \sum\nolimits_j \alpha_{ij}x_j\mathrm{d}N_j \quad (6)$$

where $u_i(x_i, t)$ determines how fast the opinion needs to be changed for user $i$. For example, a network moderator may request the user to change his opinion from $-3$ to $1$ in one day, and this control policy quantifies the amount of change in unit time.

Next we present the objective for optimizing the SDE.

**Definition 3** (Stochastic User Activity Guiding Problem)**.** *For the SDE in (5), given the initial condition $(\boldsymbol{x}_0, t_0)$, we aim to find an optimal policy $\boldsymbol{u}^*(\boldsymbol{x}, t)$ for $t \in (t_0, T]$, which minimizes the convex objective function:*

$$V(\boldsymbol{x}_0, t_0) := \min_{\boldsymbol{u}} \mathbb{E}\Big[\phi\big(\boldsymbol{x}(T)\big) + \int_{t_0}^{T} \mathcal{L}(\boldsymbol{x}, \boldsymbol{u}, t)\mathrm{d}t\Big] \quad (7)$$

*where $V$ is called the value function that summarizes the optimal expected cost if $\boldsymbol{u}^*$ is executed from $t_0$ to $T$. It is a function of the initial state. The expectation $\mathbb{E}$ is over stochastic processes $\{\boldsymbol{w}(t), \boldsymbol{N}(t)\}$ for $t \in (t_0, T]$. $\phi$ is the terminal cost and $\mathcal{L}$ is the instantaneous cost.*

**Terminal cost $\phi$.** It is the cost at final time $T$. We discuss several functional forms as follows:

- **Least square guiding.** The goal is to guide the expected state to a pre-specified target $\boldsymbol{a}$. For opinion dynamics, the goal can be to ensure nobody believes the rumor. Mathematically, we set $\phi = \|\boldsymbol{x}(T) - \boldsymbol{a}\|^2$. Moreover, to influence users' intensity of generating events, one can set the desired level of the intensity to be at a high level $\boldsymbol{a}$ and conduct activity guiding: $\phi = \|\boldsymbol{\lambda}(T) - \boldsymbol{a}\|^2$.
- **Information/activity maximization.** The goal is to maximize the activity content of all users. For example, the goal for an educator is to maximize the students' recognition of the value of education, and we set $\phi = -\sum_u x_u(T)$ to maximize each user's positive opinion. Moreover, to improve the activity level in social platforms, one can also maximize the intensity: $\phi = -\sum_u \lambda_u(T)$.

**Instantaneous cost $\mathcal{L}$.** This is the cost at $t \in [t_0, T]$ and in the form of $\mathcal{L}(\boldsymbol{x}, \boldsymbol{u}, t) = q(\boldsymbol{x}(t)) + \rho c(\boldsymbol{u}(t))$. The state cost $q(\boldsymbol{x}(t))$ is optional and the control cost $c(\boldsymbol{u}(t)) = \|\boldsymbol{u}(t)\|^2$ is necessary. We set $q = 0$ if the cost only occurs at $T$; otherwise $q$ captures the cost at the intermediate time, *e.g.*, the cost incurred by maximizing students' positive recognition of the value of education over consecutive weeks. Its function form is the same as the terminal cost: $q = \phi$. The control cost captures the fact that the policy costs money or human efforts. The scalar $\rho$ is the trade-off between control cost and state cost.

Solving this activity guiding problem is challenging, since the objective function involves taking expectation over complex stochastic processes. Furthermore, it is a functional optimization problem since the optimal policy is a function of both state and time. Fortunately, the SDE formulations allow us to connect the problem to that of stochastic dynamic programming methods. As a result, we can extend lots of tools in stochastic optimal control to address sequential decision making problems.

# 5 Algorithm for optimal policy

In this section, we will find the optimal policy posed in (7). Prior works in control theory mostly study the SDE where the jump is a Poisson process (Boel & Varaiya, 1977; Hanson, 2007; Oksendal & Sulem, 2005; Pham, 1998). However, in our model, the jump process is a more complex point process with *stochastic intensity functions*, *e.g.*, Hawkes process. Hence significant generalizations, both in theory and algorithms, are needed. We first derive the HJB partial differential equation (PDE) for a deterministic system, then generalize the procedure to the activity guiding problem. Further, we extend our framework to guide user activities in the challenging time-varying networks.

## 5.1 HJB equation for deterministic systems

To obtain the optimal policy, we need to compute the value function $V$ in (7) subject to the constraint of the SDE. We will break down the complex optimization into simpler subproblems. First, the initial condition $\boldsymbol{x}(t_0)$ needs to be replaced by an arbitrary start $\boldsymbol{x}(t)$, so that the start can be analytically manipulated and we obtain a time-varying objective $V(\boldsymbol{x}, t)$ amenable to analysis. Next, since the value function consists of an integral term, we break the integral into $[t, t + \mathrm{d}t]$ and $[t + \mathrm{d}t, T]$. If the system is deterministic, we can further split the value function as:

$$V(\boldsymbol{x}, t) = \min_{\boldsymbol{u}} \Big[ \underbrace{\phi + \int_{t+\mathrm{d}t}^{T-} \mathcal{L}\, \mathrm{d}\tau}_{V(\boldsymbol{x}(t+\mathrm{d}t), t+\mathrm{d}t)} + \underbrace{\int_{t}^{t+\mathrm{d}t} \mathcal{L}\, \mathrm{d}\tau}_{\text{cost } t \,\to\, t+\mathrm{d}t} \Big] \quad (8)$$

The first term is $V(\boldsymbol{x}(t + \mathrm{d}t), t + \mathrm{d}t)$ and the second term is the optimal cost on $[t, t + \mathrm{d}t]$. Hence (8) follows the structure of a dynamic programming and we can solve it recursively: given $V(\boldsymbol{x}(t + \mathrm{d}t), t + \mathrm{d}t)$, we only need to proceed optimally on $[t, t + \mathrm{d}t]$ to compute $V$ backward.

To further simplify (8), we perform deterministic Taylor expansion up to second order of the first term on right-hand side as $V(\boldsymbol{x}(t + \mathrm{d}t), t + \mathrm{d}t) := V(\boldsymbol{x}, t) + \mathrm{d}V(\boldsymbol{x}, t)$, where $\mathrm{d}V = V_t \mathrm{d}t + V_x^\top \mathrm{d}\boldsymbol{x} + \frac{1}{2}\mathrm{d}\boldsymbol{x}^\top V_{xx} \mathrm{d}\boldsymbol{x} + \mathrm{d}\boldsymbol{x}^\top V_{xt} \mathrm{d}t + \frac{1}{2} V_{tt} \mathrm{d}t^2$. Then we can cancel $V(\boldsymbol{x}, t)$ on both sides of (8), divide it by $\mathrm{d}t$, and take the limit as $\mathrm{d}t \to 0$. Since $\mathrm{d}\boldsymbol{x} = \boldsymbol{x}'(t)\mathrm{d}t$, all the second order term in $\mathrm{d}V$ goes to 0. Hence we obtain the Hamilton-Jacobi-Bellman (HJB) equation: $-V_t = \min_u [\mathcal{L}(\boldsymbol{x}, \boldsymbol{u}, t) + V_x^\top \boldsymbol{x}']$. However, our system is stochastic and the above procedure needs to be generalized significantly.

## 5.2 HJB equation for guiding user activities

To derive the HJB equation for our model, we need to address two challenges: (i) computing the stochastic Taylor expansion $\mathrm{d}V$ under our SDE, and (ii) taking expectation of the stochastic terms in (8) before minimization. We first derive Theorem 4 to compute $\mathrm{d}V$, which generalizes Ito's Lemma and is applicable to SDEs driven by point processes with stochastic intensity functions. Then we compute the expectation and derive a HJB equation in Theorem 5.

**Theorem 4** (Generalized Ito's Lemma for Jump Diffusion SDEs)**.** *Given the SDE in (5), let $V(\boldsymbol{x}, t)$ be twice differentiable in $\boldsymbol{x}$ and once in $t$, then we have*

$$\begin{aligned} \mathrm{d}V = &\Big\{ V_t + \frac{1}{2}\mathrm{tr}(V_{\boldsymbol{xx}} \boldsymbol{g}\boldsymbol{g}^\top) + V_{\boldsymbol{x}}^\top (\boldsymbol{f} + \boldsymbol{u}) \Big\} \mathrm{d}t + V_{\boldsymbol{x}}^\top \boldsymbol{g} \mathrm{d}\boldsymbol{w} \\ &+ \big(V(\boldsymbol{x} + \boldsymbol{h}, t) - V(\boldsymbol{x}, t)\big)\mathrm{d}\boldsymbol{N}(t) \end{aligned} \quad (9)$$

**Theorem 5** (HJB Equation)**.** *Let $\boldsymbol{h}_j(\boldsymbol{x}, t)$ be the $j$-th column of $\boldsymbol{h}(\boldsymbol{x}, t)$. Then the HJB Partial Differential Equation for the user activity guiding problem is*

$$\begin{aligned} -V_t = \min_{\boldsymbol{u}} \Big[ &\mathcal{L} + \frac{1}{2}\mathrm{tr}\big(V_{\boldsymbol{xx}} \boldsymbol{g}\boldsymbol{g}^\top\big) + V_{\boldsymbol{x}}^\top (\boldsymbol{f} + \boldsymbol{u}) \quad (10) \\ &+ \sum\nolimits_j \lambda_j(t) \big(V(\boldsymbol{x} + \boldsymbol{h}_j(\boldsymbol{x}, t), t) - V(\boldsymbol{x}, t)\big) \Big] \end{aligned}$$

To prove Theorem 4, we derive a new set of stochastic calculus rules for general point processes and conduct stochastic Taylor expansion in the Ito's mean square limit sense. Appendix C contains details. To prove Theorem 5, we combine the Generalized Ito's Lemma with the property of point processes. Appendix D contains details. Next, we solve the HJB equation to obtain the value function $V$. We will show that under the optimal parameterizations of $V$, this HJB equation can be solved efficiently.

## 5.3 Parameterization of the value function

To solve the HJB equation, we first show the structure of the value function in the following proposition.

**Proposition 6.** *If the SDE in (5) is linear in $\boldsymbol{x}$, and the terminal and instantaneous cost are quadratic/linear in $\boldsymbol{x}$, then the value function $V(\boldsymbol{x}, t)$ in (7) must be quadratic/linear.*

This result is intuitive since the $V$ is the optimal value of the summation of quadratic/linear functions. Appendix E contains the proof. This proposition is applicable to two important problems, including the least square activity guiding problem and the linear activity maximization problem.

In this section, we present derivations for the least square guiding problem, and Appendix H contains derivations for the activity maximization problem. Specifically, we set $V(\boldsymbol{x}, t)$ to be quadratic in $\boldsymbol{x}$ with unknown coefficients $\boldsymbol{v}_1(t) \in \Re^U$, $\boldsymbol{v}_{11}(t) \in \Re^{U \times U}$ and $v_0(t) \in \Re$:

$$V(\boldsymbol{x}, t) = v_0(t) + \boldsymbol{v}_1(t)^\top \boldsymbol{x} + \frac{1}{2}\boldsymbol{x}^\top \boldsymbol{v}_{11}(t) \boldsymbol{x}$$

**Algorithm 1** OPTIMAL CONTROL POLICY

**Input:** network $\boldsymbol{A} = (\alpha_{ij})$, model parameters $\{\boldsymbol{\eta}, \theta, \boldsymbol{b}\}$, timestamps $\{\tau_k\}_{k=1}^m$, events $\{t_i\}_{i=1}^n$, target $\boldsymbol{a}$
**Output:** $\boldsymbol{v}_{11}(\tau_k), \boldsymbol{v}_1(\tau_k), k = 1, \cdots, m$
**for** $k = 1$ **to** $m$ **do**
  Compute $\boldsymbol{\lambda}_u(\tau_k) = \eta_u + \sum_{j:t_i<\tau_k} \alpha_{uu_i}\kappa(\tau_k - t_i)$, $\boldsymbol{\Lambda}(\tau_k) = \sum_{j=1}^U \lambda_j(\tau_k)\boldsymbol{B}^j$ for each user $u$
**end for**
Compute $\boldsymbol{v}_{11}(\tau_k), \boldsymbol{v}_1(\tau_k)$ using ODE45; then compute $\boldsymbol{u}(\tau_k)$ in (11).

To find the optimal control, we substitute $V(\boldsymbol{x}, t)$ to the HJB equation in (10) and set the gradient of its right-hand-side to $\boldsymbol{0}$. This yields the optimal feedback control policy:

$$\boldsymbol{u}^*(\boldsymbol{x}(t), t) = -V_{\boldsymbol{x}}/\rho = -\big(\boldsymbol{v}_1(t) + \boldsymbol{v}_{11}(t)\boldsymbol{x}(t)\big)/\rho \quad (11)$$

This policy consists of two terms: the feedforward term $\boldsymbol{v}_1(t)$ controls the system as time goes by; the feedback term updates the policy based on the current state $\boldsymbol{x}(t)$. Moreover, $\rho$ controls the tradeoff between control and state cost, and $\rho \to \infty$ means low budget; hence $\boldsymbol{u}^* \to \boldsymbol{0}$.

### 5.4 Stochastic optimal control algorithm

Given the form of optimal policy $\boldsymbol{u}^*$ in (11), the final step is to compute its unknown coefficients $\{\boldsymbol{v}_1(t), \boldsymbol{v}_{11}(t)\}$. We substitute this optimal policy to the HJB equation in (10). Since the value function $V(\boldsymbol{x}, t)$ is quadratic, we can separate the HJB equation into terms that are scalar, linear, and quadratic in $\boldsymbol{x}$. Grouping coefficients of these terms leads to two Ordinary Differential Equations (ODEs) as follows:

$$-\boldsymbol{v}_{11}' = \boldsymbol{I} + 2\boldsymbol{v}_{11}(-1 + \boldsymbol{\Lambda}) - \frac{\boldsymbol{v}_{11}^2}{\rho} + \sum\nolimits_j \lambda_j \boldsymbol{B}^{j\top}\boldsymbol{v}_{11}\boldsymbol{B}^j$$

$$-\boldsymbol{v}_1'(t) = -\boldsymbol{a} + \big(-1 + \boldsymbol{\Lambda}^\top - \boldsymbol{v}_{11}(t)/\rho\big)\boldsymbol{v}_1(t) + \boldsymbol{v}_{11}(t)\boldsymbol{b}$$

where $\boldsymbol{\Lambda}(t) = \sum_j \lambda_j(t)\boldsymbol{B}^j$, matrix $\boldsymbol{B}^j$ has the $j$-th column as $(\alpha_{1j}, \cdots, \alpha_{Uj})^\top$ and zero elsewhere. The terminal conditions are $\boldsymbol{v}_{11}(T) = \boldsymbol{I}$ and $\boldsymbol{v}_1(T) = -\boldsymbol{a}$. We use the Runge-Kutta method (Dormand & Prince, 1980) to solve the ODEs offline. Specifically, we partition $(t_0, T]$ to equally-spaced timestamps $\{\tau_k\}$ and obtain values of $\boldsymbol{v}_{11}(t), \boldsymbol{v}_1(t)$ at these timestamps. We use the ODE45 solver in MATLAB. Finally we update the policy online according to (11). Algorithm 1 summarizes the procedure.

### 5.5 Extensions to time-varying networks

We can extend our framework to networks with time-varying edges and node birth processes. Specifically, for a fixed network, the expectation in the objective function (7) is over Wiener processes and point processes. In stochastic networks, the network itself adds an extra stochasticity to the SDE and we need to take the expectation of the network topology $\boldsymbol{A}(t)$ to derive the HJB equation. Hence the input to Algorithm 1 is $\mathbb{E}[\boldsymbol{A}(t)]$ instead of $\boldsymbol{A}$. Appendix B contains details on computing this expectation.

## 6 Experiments

We focus on two tasks: least square opinion guiding (LSOG) and opinion influence maximization (OIM). We compare with the following state-of-art stochastic optimization methods that are applicable to control SDEs and point processes.

- **Value Iteration (VI)** (Bertsekas, 2012): we directly formulate a discrete-time version of the opinion dynamics and compute the optimal policy using the value iteration algorithm. The discretized timestamps are the same as that for solving the HJB PDE.
- **Cross Entropy (CE)** (Stulp & Sigaud, 2012): it samples policies from a normal distribution, sorts them in ascending order w.r.t. the cost and recomputes the distribution parameters from the first K elite samples. This procedure repeats with new distribution until costs converge.
- **Finite Difference (FD)** (Peters & Schaal, 2006): it generates samples of perturbed policies and compute perturbed costs. Then it uses them to approximate the gradient of the cost w.r.t the policy. The cost in CE and FD is evaluated by executing the policy on the SDE.
- **Greedy**: It controls the SDE when the state cost is high. We divide time horizon into $n$ timestamps. At each timestamp, we compute the state cost and control the system based on pre-specified rules if current cost is more than $k$ times of the optimal cost of our method. We vary $k$ from 1 to 5, $n$ from 1 to 100 and report the best performance.
- **Slant** (De et al., 2015): It sets the open loop control policy only at the initial time.

### 6.1 Experiments on synthetic data

We generate a synthetic a network with 1000 users, where the topology matrix is randomly generated with a sparsity of 0.001. We simulate the opinion SDE on $[0, 10]$ using the Euler method (Hanson, 2007) to compute its difference form. The time window is divided into 100 equally spaced intervals. We set the base opinion uniformly at random, $b_i \sim \mathcal{U}[-1, 1]$, $\omega = 1$, noise level $\theta = 0.2$, $\alpha_{ij} \sim \mathcal{U}[0, 0.01]$, and $x_i(0) = -10$. The Wiener process is simulated from the Gaussian distribution and the Hawkes process is simulated using the Thinning algorithm (Ogata, 1981). We set the budget level parameter $\rho = 10$, and our results generalize beyond this value. We repeat simulations of the SDE for ten times and report the averaged performance. Appendix F contains details on this experimental setup.

**Total cost.** For LSOG, we set the target $a_i = 1$. The total cost per user is computed by dividing the value function by # users. Since OIM aims to maximize positive opinions, the total opinion per user is computed by dividing the negative value function by # users. Figure 2(a) shows that our method has around 2.5× improvement over CE and 4× improvement over FD. CE assumes the policy is sampled from a Gaussian distribution, and FD approximates the gradient. However, our method does not have such restrictions or approximations, and it exactly minimizes the cost.

**Importance of the SDE formulation**. Our framework significantly outperforms VI, since our SDE reformulation of user activity models preserves the *continuous time* property and our policy exactly optimizes the objective function using the HJB equation. In contrast, discretizing the original opinion model introduces approximation errors, which further influences the policies in VI. Hence it is important to directly reformulating these user activity models into SDEs and study the *continuous time* control problem.

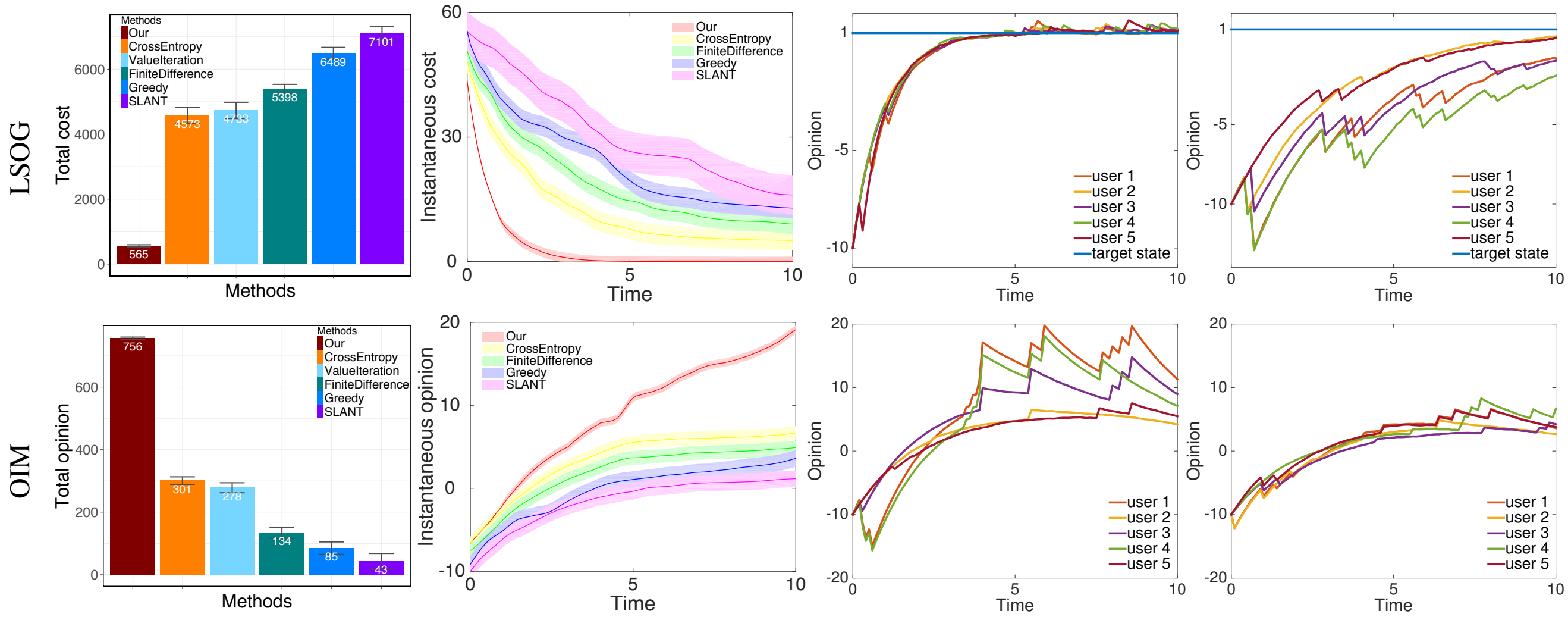

(a) Total cost/opinion (b) Cost/opinion vs. time (c) Trajectory of our work (d) Trajectory of CE

Figure 2: Results in least square guiding (LSOG) and opinion maximization (OIM). (a) total cost for LSOG and total opinion for OIM per user. Error bar is the variance; (b) instantaneous cost/opinion per user. Line is the mean and pale region is the variance; (c,d) sample trajectories of five users.

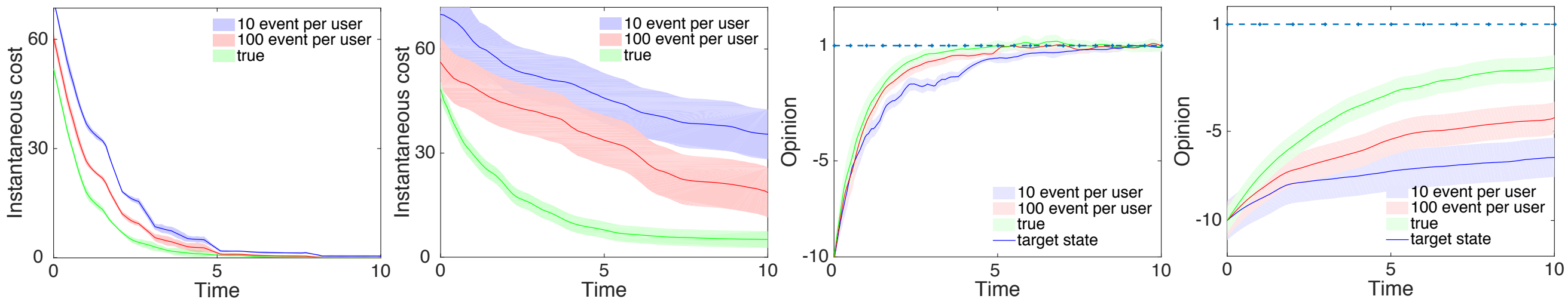

(a) Cost of our work (b) Cost of CE (c) Trajectory of our work (d) Trajectory of CE

Figure 3: Robustness analysis when parameters are learned with different sizes of training data. (a) and (b) instantaneous cost; (c) and (d) opinion trajectory for one randomly sampled user.

**Instantaneous cost & trajectory.** Figure 2(b) shows the instantaneous cost per user over time. Our method has the fastest convergence rate to the optimal cost and the cost is much smaller than competitors. Moreover, our method has the smallest variance and is stable despite multiple runs and the stochasticity in the SDE. Figure 2(c,d) compare opinion trajectories. The jumps in the opinion trajectories correspond to the opinion posting events that are modulated by the Hawkes process. For LSOG, the opinion converges to the target the fastest. For OIM, our method maximizes the opinion from the negative initial value quickly: around time 2.5, all users' opinion are positive. Moreover, our method achieves the largest opinion value, *e.g.*, 20, while that of CrossEntropy is smaller than 10.

**Robustness.** Since error exists between estimated parameters and ground truth, we investigate how our method performs with this discrepancy. We generate data with 10 and 100 events per user, and learn parameters by maximum likelihood estimation (Iacus, 2009). Figure 3(a,c) show that learned parameters are close to ground truth as the training data increases. Moreover, even with less accurate parameters, our cost and trajectories are close to ground-truth, while CE has high variance.

### 6.2 Experiments on real-world data

We study the least square opinion guiding problem over two node birth networks. *Twitter* (Farajtabar et al., 2015) contains nearly 550,000 tweet, retweet and link creation events from around 280,000 users. We use events from Sep. 21-30, 2012 and use the data before Sep. 21 to construct the initial social network. We consider the links created in the second 10-day period to be the node birth. *MemeTracker* (Leskovec et al., 2009) contains online social media activities from August 2008 to April 2009. Users track the posts of others and the network growth is captured by hyperlinks of comments on one site to others. We extract 11,321,362 posts among 5000 nodes. We use the data in Aug. 2008 to construct the initial network and use the LIWC (Pennebaker et al., 2015) toolbox to extract opinions from posts. We learn parameters of the opinion dynamics and the link creation process by maximizing the likelihood (Iacus, 2009).

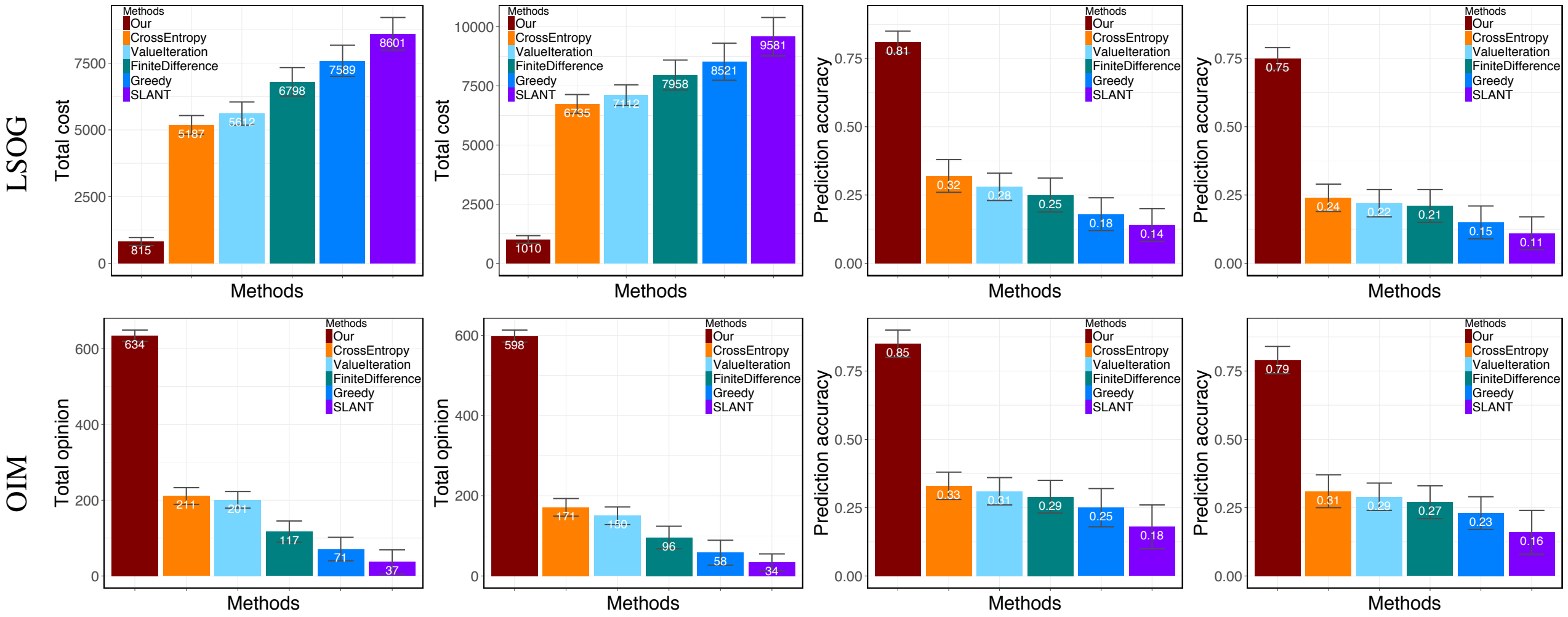

(a) Cost/opinion, Meme (b) Cost/opinion, Twitter (c) Prediction, Meme (d) Prediction, Twitter

Figure 4: Results in LSOG and OIM over real-world networks with node birth processes. (a,b) total cost (for LSOG) and opinion (for OIM) in two datasets; (c,d) prediction accuracy in two datasets.

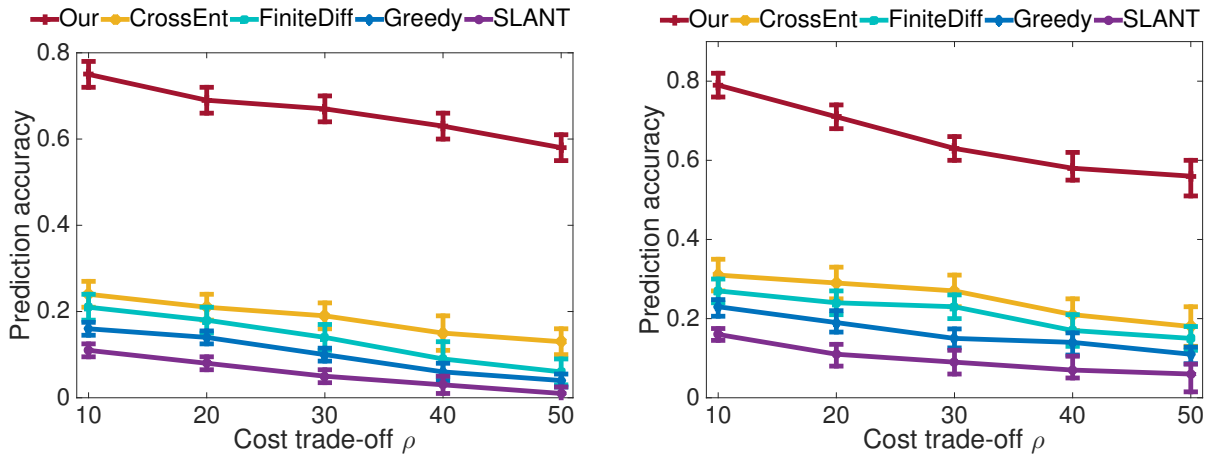

(a) Prediction in LSOG task (b) Prediction in OIM task

Figure 5: Budget sensitivity analysis: prediction accuracy as a function of $\rho$ on Twitter.

We use two evaluation procedures. First, we have a real network and learned parameters; hence we simulate user behaviors, control simulated behaviors, and evaluate the total cost of different policies. The second and more interesting evaluation scheme would entail carrying real policy in a social platform. Since it is very challenging to evaluate on a real platform, we mimic such procedure using held-out data. The key idea is to predict which real trajectory reaches the objective better (has lower cost), by comparing it to the optimal trajectory $\boldsymbol{x}^*$. Different methods yield different $\boldsymbol{x}^*$, and the prediction accuracy depends on how optimal $\boldsymbol{x}^*$ is. If it is optimal, it is accurate if we use it to order the real trajectories, and the predicted list should be similar to the ground truth, which is close to the accuracy of 1.

**Total cost.** Figure 4(a,b) show that our method performs the best for the two time-varying networks. Compared with CrossEntropy, it achieves around 6$\times$ improvement on LSOG and 3$\times$ on OIM. This result suggests that controlling the SDE over time-varying networks is a challenging problem for traditional stochastic optimal control algorithms. Moreover, the total costs of all methods for *Twitter* are higher than that of *Memetracker*. This is because *Twitter* has a much higher frequency of node birth, *i.e.*, users join the network in the timescale of minute-to-minute rather than day-to-day in *Memetracker*. Hence it is more challenging to control due to the high stochasticity in the network.

**Prediction accuracy & budget sensitivity.** Figure 4(c,d) show that our method achieves more than 0.4+ improvement over CrossEntropy. It means that our method accommodates 40% more of the total realizations correctly. An accurate prediction means that if applying our control policy, we will achieve the objective better than alternatives. Figure 5 shows that our method performs the best as the budget level decreases. Large value of the cost tradeoff parameter $\rho$ means small budget.

## 7 Conclusions

We have proposed a novel SDE reformulation for user activity models and presented the activity guiding problem, which builds a new bridge between the problem of guiding of user activities in "closed loop" and stochastic optimal control theory. Moreover, we have shown that it is important to incorporate the system status information to design a feedback control policy, which will achieve a lower cost with faster speed. Our method also provides an efficient way to guide user activities over time-varying networks with link creation and node birth processes. There are many extensions we can imagine along this direction. For example, the jump process in the SDE model can be extended to marked point processes (Jacobsen, 2006), and we can also use nonlinear SDEs to model user activities.

**Acknowledgements**. This project was supported in part by NSF IIS-1218749, NIH BIGDATA 1R01GM108341, NSF CAREER IIS-1350983, NSF IIS-1639792 EAGER, NSF CNS-1704701, ONR N00014-15-1-2340, DMS-1620342, CMMI-1745382, IIS-1639792, IIS-1717916, NVIDIA, Intel ISTC and Amazon AWS.

# A Proof of Theorem 2

Given any two function $f(t)$ and $g(t)$, we first define a convolution operator $\star$ as follows

$$f(t) \star g(t) = \int_0^t f(t-s)g(s)\mathrm{d}s \tag{12}$$

Therefore, the user activity model for $x_i(t)$ can be expressed as

$$x_i(t) = \underset{\substack{\uparrow \\ \text{base}}}{b_i(t)} + \underbrace{\sum_{j=1}^{U} \alpha_{ij}\kappa_{\omega_2}(t) \star \big(h(x_j(t))\mathrm{d}N_j(t)\big)}_{\text{social neighbor influence}} \tag{13}$$

Before we apply the differential operator $\mathrm{d}$ to (13), we also need the following two properties:

- $\mathrm{d}\kappa_{\omega_2}(t) = -\omega_2\kappa_{\omega_2}(t)\mathrm{d}t$ for $t \geqslant 0$ and $\kappa_{\omega_2}(0) = 1$.
- The differential of the convolution of two functions is expressed as: $\mathrm{d}(f \star g) = f(0)g + g \star \mathrm{d}f$.

With the above two properties, we set $f = \kappa_{\omega_2}(t)$ and $g = \sum_j \alpha_{ij}h(x_j)\mathrm{d}N_j(t)$, and take the differential of $x_i(t)$ in (13) as follows

$$\begin{aligned}
\mathrm{d}x_i(t) &= \mathrm{d}b_i(t) + \mathrm{d}(f \star g) &(14)\\
&= \mathrm{d}b_i(t) + \sum_{j=1}^{U} \alpha_{ij}h(x_j)\mathrm{d}N_j(t) - \omega_2\Big(\sum_{j=1}^{U} \alpha_{ij}k_{\omega_2}(t) \star (h(x_j)\cdot \mathrm{d}N_j(t))\Big)\mathrm{d}t &(15)\\
&= \mathrm{d}b_i(t) + \sum_{j=1}^{U} \alpha_{ij}h(x_j)\mathrm{d}N_j(t) - \omega_2(x_i(t) - b_i(t))\mathrm{d}t &(16)\\
&= \mathrm{d}b_i(t) + \omega_2(b_i(t) - x_i(t))\mathrm{d}t + \sum_{j=1}^{U} \alpha_{ij}h(x_j(t))\mathrm{d}N_j(t) &(17)
\end{aligned}$$

This completes the proof for the SDE formulation of $x_i(t)$.

Similarly, we can express the intensity function using the convolution operator as follows

$$\lambda_i(t) = \underset{\substack{\uparrow \\ \text{base}}}{\eta_i(t)} + \underbrace{\sum_{j=1}^{U} \beta_{ij}\kappa_{\omega_1}(t) \star \mathrm{d}N_j(t)}_{\text{social neighbor influence}} \tag{18}$$

Then we set $f = \kappa_{\omega_1}(t)$, $g = \sum_j \beta_{ij}\mathrm{d}N_j(t)$, and can show the following equation:

$$\mathrm{d}\lambda_i(t) = \mathrm{d}\eta_i(t) + \omega_1\big(\eta_i(t) - \lambda_i(t)\big)\mathrm{d}t + \sum_j \beta_{ij}\mathrm{d}N_j(t) \tag{19}$$

This completes the proof.

## B Extensions to time-varying networks

Real world social networks can change over time. Users can follow or unfollow each other as time goes by and new users can join the network (Farajtabar et al., 2015). In this section, we extend our framework to networks with time-varying edges and node birth processes.

First, for a fixed network, the expectation in the objective function in (7) is over the stochastic pair $\{\boldsymbol{w}(t), \boldsymbol{N}(t)\}$ for $t \in (t_0, T]$. Since the network is stochastic now, we also need to take the expectation of the adjacency matrix $\boldsymbol{A}(t) = (\alpha_{ij}(t))$ to derive the HJB equation. Hence the input to Algorithm 1 is $\mathbb{E}[\boldsymbol{A}(t)] = (\mathbb{E}[\alpha_{ij}(t)])$ instead of $\boldsymbol{A}$. Specifically, we replace $\boldsymbol{h}_j(\boldsymbol{x})$ in the HJB equation (10) by $\mathbb{E}[\boldsymbol{h}_j(\boldsymbol{x})]$:

$$\sum\nolimits_j \lambda_j(t)(V(\boldsymbol{x} + \mathbb{E}[\boldsymbol{h}_j(\boldsymbol{x}, t)], t) - V(\boldsymbol{x}, t)) \tag{20}$$

where $\mathbb{E}[\boldsymbol{h}_j(\boldsymbol{x}, t)] = (\mathbb{E}[h_{1j}(t)], \cdots, \mathbb{E}[h_{Uj}(t)])^\top$ and $\mathbb{E}[h_{ij}(t)] = \mathbb{E}[\alpha_{ij}(t)]x_j(t)$. Next, we compute $\mathbb{E}[\alpha_{ij}(t)]$ in two types of networks.

**Networks with link creation**. We model the creation of link from node $i \to j$ as a survival process $\alpha_{ij}(t)$. If a link is created, $\alpha_{ij}(t) = 1$ and zero otherwise. Its intensity function is defined as

$$\sigma_{ij}(t) = (1 - \alpha_{ij}(t))\gamma_i, \tag{21}$$

where the term $\gamma_i \geqslant 0$ denotes the Poisson intensity, which models the node $i$'s own initiative to create links to others. The coefficient $1 - \alpha_{ij}(t)$ ensures a link is created only once, and intensity is set to 0 after that. Given a sequence of link creation events, we can learn $\{\gamma_i\}$ using maximum likelihood estimation (Aalen et al., 2008) as follows.

**Parameter estimation of the link creation process**. Given data $e_i = (t_i, u_i, s_i)$, which means at time $t_i$ node $u_i$ is added to the network and connects to $s_i$, we set $\mathcal{E} = \{e_i\}$ and optimize the concave log-likelihood function to learn the parameters of the Poisson intensity $\boldsymbol{\gamma} = (\gamma_1, \cdots, \gamma_U)^\top$:

$$\max_{\boldsymbol{\gamma} \geqslant 0} \sum_{e_i \in \mathcal{E}} \log(\sigma_{u_i s_i}(t_i)) - \sum_{u,s \in [n]} \int_0^T \sigma_{us}(\tau) \mathrm{d}\tau$$

This objective function can be solved efficiently with many optimization algorithms, such as the Quasi-Newton algorithm.

Next, given the learned parameters, we obtain the following ordinary differential equation (ODE) that describes the time-evolution of $\mathbb{E}[\alpha_{ij}(t)]$:

$$\mathrm{d}\mathbb{E}[\alpha_{ij}(t)] \underset{(a)}{=} \mathbb{E}[\mathrm{d}\alpha_{ij}(t)] \underset{(b)}{=} \sigma_{ij}(t)\mathrm{d}t \underset{(c)}{=} (1 - \mathbb{E}[\alpha_{ij}(t)])\gamma_i \mathrm{d}t, \tag{22}$$

where (a) holds because the operator d and $\mathbb{E}$ are exchangeable, (b) is from the definition of intensity function, and (c) is from (21). The initial condition is $\mathbb{E}[\alpha_{ij}(0)] = 0$ since $i$ and $j$ are not connected initially. We can easily solve this ODE in analytical form:

$$\mathbb{E}[\alpha_{ij}(t)] = 1 - \exp(-\gamma_i t)$$

**Networks with node birth**. The network's dimension can grow as new users join it. Since the dimension of $\boldsymbol{A}(t)$ changes over time, it is very challenging to control such network, and it remains unknown how to derive the HJB equation for such case. We propose an efficient method by connecting the stochasticity of the node birth process to that of link creation process. More specifically, we have the following observation.

**Observation.** *The process of adding a new user $v$ to the existing network $\boldsymbol{A} \in \Re^{(N-1)\times(N-1)}$ and connects to user $s$ is equivalent to link creation process of setting $\boldsymbol{A}(t) \in \Re^{N \times N}$ to be the existing network and letting $\alpha_{vs}(t) = 1$.*

With this observation, we can fix the dimension of $\boldsymbol{A}(t)$ beforehand, and add a link whenever a user joins the network. This procedure is memory-efficient since we do not need to maintain a sequence of size-growing matrices. More importantly, we transform the stochasticity of the network's dimension to the stochasticity of link creation process with a fixed network dimension. Finally, the difference between link creation and node birth is: we control each node in the link creation case, but do not control the node until it joins the network in the node birth case.

# C Proof of Theorem 4

**Theorem 4** (Generalized Ito's Lemma)**.** *Given the SDE in (5), let $V(\boldsymbol{x}, t)$ be twice-differentiable in $\boldsymbol{x}$ and once in $t$; then we have:*

$$\mathrm{d}V = \Big\{V_t + \frac{1}{2}\mathrm{tr}(V_{\boldsymbol{xx}}\boldsymbol{g}\boldsymbol{g}^\top) + V_{\boldsymbol{x}}^\top(\boldsymbol{f} + \boldsymbol{u})\Big\}\mathrm{d}t + V_{\boldsymbol{x}}^\top \boldsymbol{g}\mathrm{d}\boldsymbol{w} + \big(V(\boldsymbol{x} + \boldsymbol{h}, t) - V(\boldsymbol{x}, t)\big)\mathrm{d}\boldsymbol{N}(t) \tag{23}$$

To prove the theorem, we will first provide some background and useful formulas as follows.

$$(\mathrm{d}t)^2 = 0, \mathrm{d}t\mathrm{d}\boldsymbol{N}(t) = 0, \mathrm{d}t\mathrm{d}\boldsymbol{w}(t) = 0, \mathrm{d}\boldsymbol{w}(t)\mathrm{d}\boldsymbol{N}(t) = 0, \mathrm{d}\boldsymbol{w}(t)\mathrm{d}\boldsymbol{w}(t)^\top = \mathrm{d}t\boldsymbol{I} \tag{24}$$

All the above equations hold in the *mean square limit* sense. The mean square limit definition enables us to extend the calculus rules for deterministic functions and properly define stochastic calculus rules such as stochastic differential and stochastic integration for stochastic processes. See (Hanson, 2007) for the proof of these equations.

**Proof.** We first restate the SDE in (5) as follows

$$\begin{aligned}\mathrm{d}\boldsymbol{x} &= \big(\boldsymbol{f}(\boldsymbol{x}) + \boldsymbol{u}\big)\mathrm{d}t + \boldsymbol{g}(\boldsymbol{x})\mathrm{d}\boldsymbol{w}(t) + \boldsymbol{h}(\boldsymbol{x})\mathrm{d}\boldsymbol{N}(t)\\ &= \boldsymbol{F}(\boldsymbol{x}) + \boldsymbol{h}(\boldsymbol{x})\mathrm{d}\boldsymbol{N}(t),\end{aligned}$$

where $\boldsymbol{F}(\boldsymbol{x})$ denotes the continuous part of the SDE and is define as

$$\boldsymbol{F}(\boldsymbol{x}) = \big(\boldsymbol{f}(\boldsymbol{x}) + \boldsymbol{u}\big)\mathrm{d}t + \boldsymbol{g}(\boldsymbol{x})\mathrm{d}\boldsymbol{w}(t)$$

Note that the term $\boldsymbol{h}\mathrm{d}\boldsymbol{N}(t)$ denotes the discontinuous part of the SDE. For the simplicity of notation, we set $\boldsymbol{F}(\boldsymbol{x}) = \boldsymbol{F}$ and $\boldsymbol{h}(\boldsymbol{x}) = \boldsymbol{h}$ and omit the dependency on $\boldsymbol{x}$.

Next, we expand $\mathrm{d}V$ according to its definition as follows

$$\mathrm{d}V(\boldsymbol{x}, t) = V(\boldsymbol{x}(t + \mathrm{d}t), t + \mathrm{d}t) - V(\boldsymbol{x}, t)$$

With the definition $\boldsymbol{x}(t + \mathrm{d}t) = \boldsymbol{x}(t) + \mathrm{d}\boldsymbol{x}$, we can expand $V(\boldsymbol{x}(t + \mathrm{d}t), t + \mathrm{d}t)$ using Taylor expansion on variable $t$ as follows

$$V(\boldsymbol{x}(t + \mathrm{d}t), t + \mathrm{d}t) = V(\boldsymbol{x} + \mathrm{d}\boldsymbol{x}, t + \mathrm{d}t) \tag{25}$$

$$= V(\boldsymbol{x} + \mathrm{d}\boldsymbol{x}, t) + V_t(\boldsymbol{x}, t)\mathrm{d}t \tag{26}$$

Next, we expand $V(\boldsymbol{x} + \mathrm{d}\boldsymbol{x}, t)$ as follows

$$V(\boldsymbol{x} + \mathrm{d}\boldsymbol{x}, t)$$

$$= V(\boldsymbol{x} + \boldsymbol{F} + \boldsymbol{h}\mathrm{d}\boldsymbol{N}(t), t) \tag{27}$$

$$= \Big(V(\boldsymbol{x} + \boldsymbol{F} + \boldsymbol{h}, t) - V(\boldsymbol{x} + \boldsymbol{F}, t)\Big)\mathrm{d}\boldsymbol{N}(t) + V(\boldsymbol{x} + \boldsymbol{F}, t) \tag{28}$$

$$= \Big[\underbrace{V(\boldsymbol{x} + \boldsymbol{h}, t) + V_{\boldsymbol{x}}(\boldsymbol{x} + \boldsymbol{h})^\top \boldsymbol{F} + \frac{1}{2}\boldsymbol{F}V_{\boldsymbol{xx}}(\boldsymbol{x} + \boldsymbol{h})\boldsymbol{F}^\top}_{\text{Taylor expansion 1}} - \Big(\underbrace{V(\boldsymbol{x}, t) + V_{\boldsymbol{x}}^\top \boldsymbol{F} + \frac{1}{2}\boldsymbol{F}V_{\boldsymbol{xx}}\boldsymbol{F}^\top}_{\text{Taylor expansion 2}}\Big)\Big]\mathrm{d}\boldsymbol{N}(t) \tag{29}$$

$$+ \underbrace{V(\boldsymbol{x}, t) + V_{\boldsymbol{x}}^\top \boldsymbol{F} + \frac{1}{2}\boldsymbol{F}V_{\boldsymbol{xx}}\boldsymbol{F}^\top}_{\text{Taylor expansion 2}} \tag{30}$$

$$= \Big(V(\boldsymbol{x} + \boldsymbol{h}, t) - V(\boldsymbol{x}, t)\Big)\mathrm{d}\boldsymbol{N}(t) + \Big(V_{\boldsymbol{x}}(\boldsymbol{x} + \boldsymbol{h}) - V_{\boldsymbol{x}}\Big)^\top \boldsymbol{F}\mathrm{d}\boldsymbol{N}(t) \tag{31}$$

$$+ V(\boldsymbol{x}, t) + V_{\boldsymbol{x}}^\top \boldsymbol{F} + \frac{1}{2}\boldsymbol{F}V_{\boldsymbol{xx}}\boldsymbol{F}^\top + \Big(\frac{1}{2}\boldsymbol{F}V_{\boldsymbol{xx}}(\boldsymbol{x} + \boldsymbol{h})\boldsymbol{F}^\top - \frac{1}{2}\boldsymbol{F}V_{\boldsymbol{xx}}(\boldsymbol{x})\boldsymbol{F}^\top\Big)\mathrm{d}\boldsymbol{N}(t) \tag{32}$$

Next, we show the reasoning from (27) to (30).

First, we derive a stochastic calculus rule for the point process. Specifically, since $\mathrm{d}\boldsymbol{N}(t) \in \{0, 1\}$, there are two cases for (27): If a jump happens, *i.e.*, $\mathrm{d}\boldsymbol{N}(t) = 1$, (27) is equivalent to $V(\boldsymbol{x} + \boldsymbol{F}(\boldsymbol{x}) + \boldsymbol{h}(\boldsymbol{x}), t)$; otherwise, we have $\mathrm{d}\boldsymbol{N}(t) = 0$ and (27) is equivalent to $V(\boldsymbol{x} + \boldsymbol{F}(\boldsymbol{x}), t)$. Hence (28) is equivalent to (27). This stochastic rule essentially takes $\mathrm{d}\boldsymbol{N}(t)$ from inside the value function $V$ to the outside.

Second, from (28) to (30), we have used the following Taylor expansions.

*Taylor expansion 1.* For $V(\boldsymbol{x}+\boldsymbol{F}+\boldsymbol{h},t)$, we expand it around $V(\boldsymbol{x}+\boldsymbol{h},t)$ on the $\boldsymbol{x}$-dimension

$$V(\boldsymbol{x}+\boldsymbol{F}+\boldsymbol{h},t)=V(\boldsymbol{x}+\boldsymbol{h},t)+V_{\boldsymbol{x}}(\boldsymbol{x}+\boldsymbol{h})^{\top}\boldsymbol{F}+\frac{1}{2}\boldsymbol{F}V_{\boldsymbol{x}\boldsymbol{x}}(\boldsymbol{x}+\boldsymbol{h})\boldsymbol{F}^{\top}$$

*Taylor expansion 2.* For $V(\boldsymbol{x}+\boldsymbol{F},t)$, we expand it around $V(\boldsymbol{x},t)$ along the $\boldsymbol{x}$ dimension

$$V(\boldsymbol{x}+\boldsymbol{F},t)=V(\boldsymbol{x},t)+V_{\boldsymbol{x}}^{\top}\boldsymbol{F}+\frac{1}{2}\boldsymbol{F}V_{\boldsymbol{x}\boldsymbol{x}}\boldsymbol{F}^{\top}$$

Next, we simplify each term in (31) and (32). We keep the first term and expand the second term, $\Big(V_{\boldsymbol{x}}(\boldsymbol{x}+\boldsymbol{h})-V_{\boldsymbol{x}}\Big)^{\top}\boldsymbol{F}\mathrm{d}\boldsymbol{N}(t)$ as follows

$$\begin{aligned}\Big(V_{\boldsymbol{x}}(\boldsymbol{x}+\boldsymbol{h})-V_{\boldsymbol{x}}\Big)^{\top}\boldsymbol{F}\mathrm{d}\boldsymbol{N}(t)&=\Big(V_{\boldsymbol{x}}(\boldsymbol{x}+\boldsymbol{h})-V_{\boldsymbol{x}}\Big)^{\top}((\boldsymbol{f}+\boldsymbol{u})\mathrm{d}t+\boldsymbol{g}\mathrm{d}\boldsymbol{w}(t))\mathrm{d}\boldsymbol{N}(t)\\&=\Big(V_{\boldsymbol{x}}(\boldsymbol{x}+\boldsymbol{h})-V_{\boldsymbol{x}}\Big)^{\top}\Big((\boldsymbol{f}+\boldsymbol{u})\mathrm{d}t\mathrm{d}\boldsymbol{N}(t)+\boldsymbol{g}\mathrm{d}\boldsymbol{w}(t)\mathrm{d}\boldsymbol{N}(t)\Big),\\&=0\end{aligned}\tag{33}$$

where we have used the equations: $\mathrm{d}t\mathrm{d}\boldsymbol{N}(t)=0$ and $\mathrm{d}\boldsymbol{w}(t)\mathrm{d}\boldsymbol{N}(t)=0$ in the Ito mean square limit sense from (24).

We keep the third term and expand the fourth term $V_{\boldsymbol{x}}^{\top}\boldsymbol{F}$ as

$$V_{\boldsymbol{x}}^{\top}\boldsymbol{F}=V_{\boldsymbol{x}}^{\top}(\boldsymbol{f}+\boldsymbol{u})\mathrm{d}t+V_{\boldsymbol{x}}^{\top}\boldsymbol{g}\mathrm{d}\boldsymbol{w}(t)\tag{34}$$

The fifth term $\frac{1}{2}\boldsymbol{F}V_{\boldsymbol{x}\boldsymbol{x}}\boldsymbol{F}^{\top}$ is expanded as follows

$$\begin{aligned}&\frac{1}{2}\boldsymbol{F}V_{\boldsymbol{x}\boldsymbol{x}}\boldsymbol{F}^{\top}\\&=\frac{1}{2}\Big((\boldsymbol{f}+\boldsymbol{u})\mathrm{d}t+\boldsymbol{g}\mathrm{d}\boldsymbol{w}(t)\Big)V_{\boldsymbol{x}\boldsymbol{x}}\Big((\boldsymbol{f}+\boldsymbol{u})\mathrm{d}t+\boldsymbol{g}\mathrm{d}\boldsymbol{w}(t)\Big)^{\top}\\&=\frac{1}{2}\Big((\boldsymbol{f}+\boldsymbol{u})V_{\boldsymbol{x}\boldsymbol{x}}(\boldsymbol{f}+\boldsymbol{u})^{\top}(\mathrm{d}t)^2+2(\boldsymbol{f}+\boldsymbol{u})\mathrm{d}tV_{\boldsymbol{x}\boldsymbol{x}}(\boldsymbol{g}\mathrm{d}\boldsymbol{w}(t))^{\top}+(\boldsymbol{g}\mathrm{d}\boldsymbol{w}(t))V_{\boldsymbol{x}\boldsymbol{x}}(\boldsymbol{g}\mathrm{d}\boldsymbol{w}(t))^{\top}\Big)\\&=\frac{1}{2}\Big(0+0+\mathrm{tr}(V_{\boldsymbol{x}\boldsymbol{x}}\boldsymbol{g}\boldsymbol{g}^{\top})\mathrm{d}t\Big)\\&=\frac{1}{2}\mathrm{tr}(V_{\boldsymbol{x}\boldsymbol{x}}\boldsymbol{g}\boldsymbol{g}^{\top})\mathrm{d}t,\end{aligned}\tag{35}$$

where we have used the property that $(\mathrm{d}t)^2=0$, $\mathrm{d}t\mathrm{d}\boldsymbol{w}=0$, and $\mathrm{d}\boldsymbol{w}(t)\mathrm{d}\boldsymbol{w}(t)^{\top}=\mathrm{d}t\boldsymbol{I}$ from (24).

Finally, the last term is expressed as

$$\begin{aligned}&\Big(\frac{1}{2}\boldsymbol{F}V_{\boldsymbol{x}\boldsymbol{x}}(\boldsymbol{x}+\boldsymbol{h})\boldsymbol{F}^{\top}-\frac{1}{2}\boldsymbol{F}V_{\boldsymbol{x}\boldsymbol{x}}(\boldsymbol{x})\boldsymbol{F}^{\top}\Big)\mathrm{d}\boldsymbol{N}(t)\\&=\frac{1}{2}\mathrm{tr}(V_{\boldsymbol{x}\boldsymbol{x}}(\boldsymbol{x}+\boldsymbol{h})\boldsymbol{g}\boldsymbol{g}^{\top})\mathrm{d}t\mathrm{d}\boldsymbol{N}(t)-\frac{1}{2}\mathrm{tr}(V_{\boldsymbol{x}\boldsymbol{x}}\boldsymbol{g}\boldsymbol{g}^{\top})\mathrm{d}t\mathrm{d}\boldsymbol{N}(t)=0-0=0\end{aligned}\tag{36}$$

Substituting equation (33), (34), (35), and (36) to equation (31) and (32), we have:

$$\begin{aligned}V(\boldsymbol{x}+\mathrm{d}\boldsymbol{x},t)=&\Big(V(\boldsymbol{x}+\boldsymbol{h},t)-V(\boldsymbol{x},t)\Big)\mathrm{d}\boldsymbol{N}(t)+V_{\boldsymbol{x}}^{\top}(\boldsymbol{f}+\boldsymbol{u})\mathrm{d}t+V_{\boldsymbol{x}}^{\top}\boldsymbol{g}\mathrm{d}\boldsymbol{w}(t)\\&+V(\boldsymbol{x},t)+\frac{1}{2}\mathrm{tr}(V_{\boldsymbol{x}\boldsymbol{x}}\boldsymbol{g}\boldsymbol{g}^{\top})\mathrm{d}t\end{aligned}\tag{37}$$

Plugging (37) to (26), we have:

$$\begin{aligned}V(\boldsymbol{x}(t+\mathrm{d}t),t+\mathrm{d}t)=&\Big(V(\boldsymbol{x}+\boldsymbol{h},t)-V(\boldsymbol{x},t)\Big)\mathrm{d}\boldsymbol{N}(t)+V_{\boldsymbol{x}}^{\top}(\boldsymbol{f}+\boldsymbol{u})\mathrm{d}t+V_{\boldsymbol{x}}^{\top}\boldsymbol{g}\mathrm{d}\boldsymbol{w}(t)\\&+V(\boldsymbol{x},t)+\frac{1}{2}\mathrm{tr}(V_{\boldsymbol{x}\boldsymbol{x}}\boldsymbol{g}\boldsymbol{g}^{\top})\mathrm{d}t+V_t(\boldsymbol{x},t)\mathrm{d}t\end{aligned}$$

Hence after simplification, we obtain (23) and finishes the proof:

$$\begin{aligned}\mathrm{d}V&=V(\boldsymbol{x}(t+\mathrm{d}t),t+\mathrm{d}t)-V(\boldsymbol{x}(t),t)\\&=\Big\{V_t+\frac{1}{2}\mathrm{tr}(V_{\boldsymbol{x}\boldsymbol{x}}\boldsymbol{g}\boldsymbol{g}^{\top})+V_{\boldsymbol{x}}^{\top}(\boldsymbol{f}+\boldsymbol{u})\Big\}\mathrm{d}t+V_{\boldsymbol{x}}^{\top}\boldsymbol{g}\mathrm{d}\boldsymbol{w}+\big(V(\boldsymbol{x}+\boldsymbol{h},t)-V(\boldsymbol{x},t)\big)\mathrm{d}\boldsymbol{N}(t)\end{aligned}$$

# D Proof of Theorem 5

**Theorem 5.** *The HJB equation for the user activity guiding problem in (7) is*

$$-V_t = \min_{\boldsymbol{u}} \Big[ \mathcal{L} + \frac{1}{2} \mathrm{tr}\big(V_{\boldsymbol{x}\boldsymbol{x}} \boldsymbol{g}\boldsymbol{g}^\top\big) + V_{\boldsymbol{x}}^\top (\boldsymbol{f} + \boldsymbol{u}) + \sum\nolimits_{j=1}^{U} \lambda_j(t) \big( V(\boldsymbol{x} + \boldsymbol{h}_j(\boldsymbol{x}), t) - V(\boldsymbol{x}, t) \big) \Big]$$

*where* $\boldsymbol{h}_j(\boldsymbol{x})$ *is the* $j$*-th column of* $\boldsymbol{h}(\boldsymbol{x})$*.*

**Proof.** First we express the value function $V$ as follows

$$V(\boldsymbol{x}, t) = \min_{\boldsymbol{u}} \mathbb{E}\Big[ V(\boldsymbol{x}(t + \mathrm{d}t), t + \mathrm{d}t) + \int_t^{t+\mathrm{d}t} \mathcal{L}\, \mathrm{d}\tau \Big] \tag{38}$$

$$= \min_{\boldsymbol{u}} \mathbb{E}\Big[ V(\boldsymbol{x}, t) + \mathrm{d}V + \mathcal{L}\, \mathrm{d}t \Big] \tag{39}$$

$$= \min_{\boldsymbol{u}} \mathbb{E}\Big[ V(\boldsymbol{x}, t) + \Big\{ V_t + \frac{1}{2} \mathrm{tr}(V_{\boldsymbol{x}\boldsymbol{x}} \boldsymbol{g}\boldsymbol{g}^\top) + V_{\boldsymbol{x}}^\top (\boldsymbol{f} + \boldsymbol{u}) \Big\} \mathrm{d}t + V_{\boldsymbol{x}}^\top \boldsymbol{g} \mathrm{d}\boldsymbol{w} + \big( V(\boldsymbol{x} + \boldsymbol{h}, t) - V(\boldsymbol{x}, t) \big) \mathrm{d}\boldsymbol{N}(t) + \mathcal{L}\, \mathrm{d}t \Big] \tag{40}$$

$$= \min_{\boldsymbol{u}} \Big[ V(\boldsymbol{x}, t) + \Big\{ V_t + \mathcal{L} + \frac{1}{2} \mathrm{tr}\big(V_{\boldsymbol{x}\boldsymbol{x}} \boldsymbol{g}\boldsymbol{g}^\top\big) + V_{\boldsymbol{x}}^\top (\boldsymbol{f} + \boldsymbol{u}) \Big\} \mathrm{d}t + \sum\nolimits_{j=1}^{U} \lambda_j(t) \big( V(\boldsymbol{x} + \boldsymbol{h}_j(\boldsymbol{x}), t) - V(\boldsymbol{x}, t) \big) \mathrm{d}t, \Big] \tag{41}$$

where (39) to (40) follows from Theorem 4, and (40) to (41) follows from the properties of Wiener processes and point processes, *i.e.*, $\mathbb{E}[\mathrm{d}\boldsymbol{w}] = 0$ and $\mathbb{E}[\mathrm{d}\boldsymbol{N}(t)] = \boldsymbol{\lambda}(t)\mathrm{d}t$.

Finally, cancelling $V(\boldsymbol{x}, t)$ on both sides of (41) and dividing both sides by $\mathrm{d}t$ yields the HJB equation.

# E Proof of Proposition 6

For the quadratic cost case (the opinion least square guiding problem), we have: $\phi = \frac{1}{2}\|\boldsymbol{x}(T) - \boldsymbol{a}\|^2$, $\mathcal{L} = \frac{1}{2}\|\boldsymbol{x}(t) - \boldsymbol{a}\|^2 + \frac{\rho}{2}\|\boldsymbol{u}(t)\|^2$. Since the instantaneous cost $\mathcal{L}$ is quadratic in $\boldsymbol{x}$ and $\boldsymbol{u}$, and terminal cost $\phi$ is quadratic in $\boldsymbol{x}$, if the control $\boldsymbol{u}$ is a linear function of $\boldsymbol{x}$, then the value function $V$ must be quadratic in $\boldsymbol{x}$, since it is the optimal value of the summation of quadratic functions.

Moreover, the fact that $\boldsymbol{u}$ is linear in $\boldsymbol{x}$ is because our SDE model for user activities is linear in both $\boldsymbol{x}$ and $\boldsymbol{u}$. Since $V(T) = \phi(T)$ is quadratic, as illustrated in (Hanson, 2007), one can show by induction that when computing the value of $V$ backward in time, $\boldsymbol{u}$ is always linear in $\boldsymbol{x}$.

Similarly, one can show that the value function $V$ is linear in the state $\boldsymbol{x}$ for the linear cost case (opinion maximization problem), where $\phi = -\sum_u x_u(T)$, $\mathcal{L} = -\sum_u x_u(t) + \frac{\rho}{2}\|\boldsymbol{u}(t)\|^2$.

# F Additional synthetic experiments

**Synthetic experimental setup.** We consider a network with $1000$ users, where the network topology matrix is randomly generated with a sparsity of $0.001$. We simulate the opinion SDE on the observation window $[0, 10]$ by applying Euler forward method to compute the difference form of (6) with $\omega = 1$:

$$x_i(t_{k+1}) = x_i(t_k) + \big(b_i + u_i(t_k) - x_i(t_k)\big)\Delta t + \theta \Delta w_i(t_k) + \sum\nolimits_{j=1}^{U} \alpha_{ij} x_j(t_k) \Delta N_j(t_k),$$

where the observation window is divided into $100$ time stamps $\{t_k\}$, with interval length $\Delta t = 0.1$. The Wiener increments $\Delta w_i$ is sampled from the normal distribution $\mathcal{N}(0, \sqrt{\Delta t})$ and the Hawkes increments $\Delta N_j(t_k)$ is computed by counting the number of events on $[t_k, t_{k+1})$ for user $j$. The events for each user is simulated by the Otaga's thinning algorithm (Ogata, 1981). The thinning algorithm is essentially a rejection sampling algorithm where samples are first proposed from a homogeneous Poisson process and then samples are kept according to the ratio between the actual intensity and that of the Poisson process.

We set the baseline opinion uniformly at random, $b_i \sim \mathcal{U}[-1, 1]$, noise level $\theta = 0.2$, $\alpha_{ij} \sim \mathcal{U}[0, 0.01]$, initial opinion $x_i(0) = -10$, and $\omega = 1$ for the exponential triggering kernel $\kappa_\omega$. We repeat simulation of the SDE for ten times and report average performance. We set the tradeoff (budget level) parameter $\rho = 10$, and our results generalize beyond this value.

**Network visualization.** We conduct control over this 1000-user network with four different initial and target states. Figure 6 shows that our framework works efficiently.

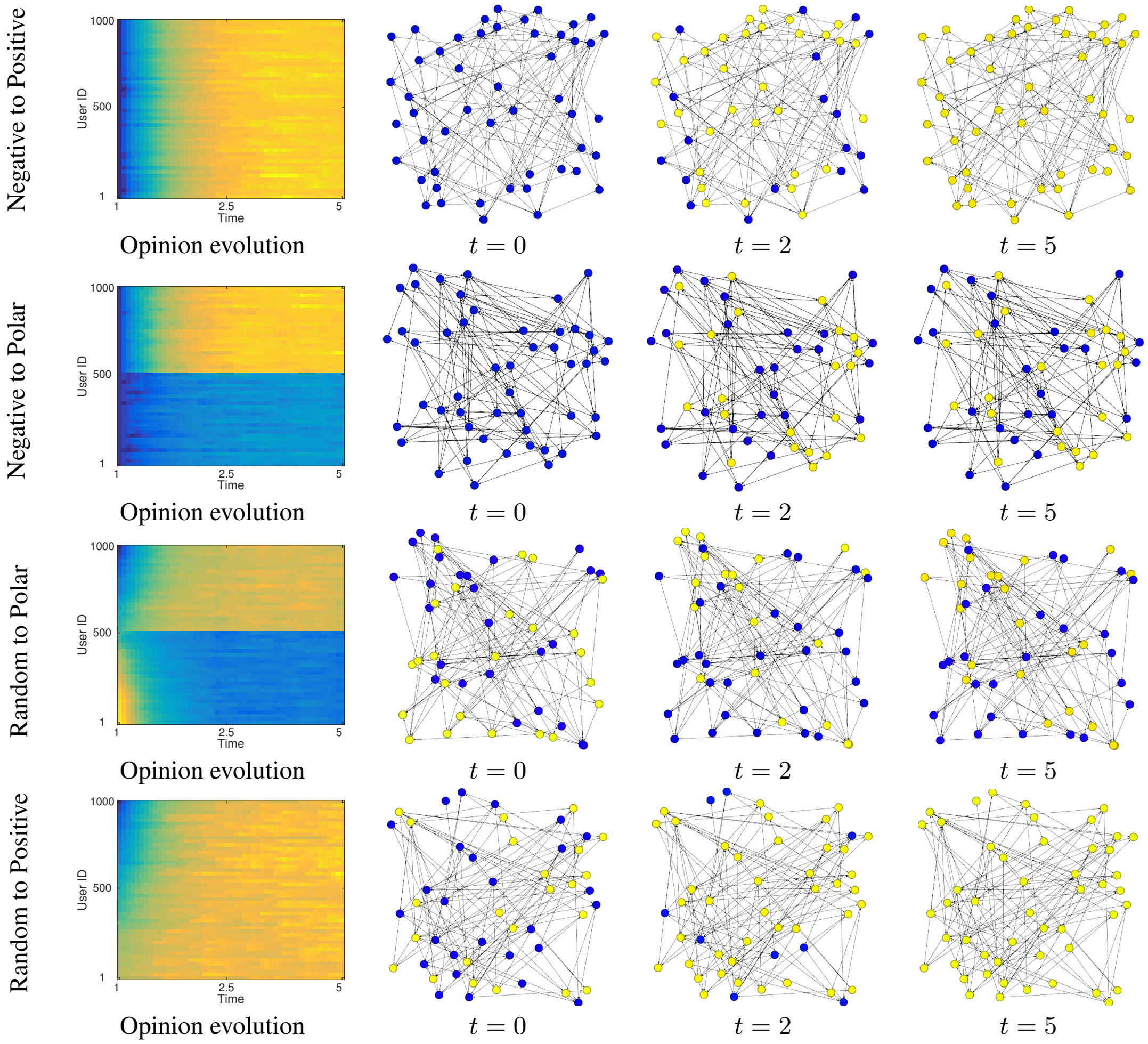

Figure 6: Controlled opinion of four networks with 1,000 users. The first column is the description of opinion change. The second column shows the opinion *value* per user over time. The three right columns show three snapshots of the opinion *polarity* in the network with 50 sub-users at different times. Yellow means positive and blue means negative polarity. Since the controlled trajectory converges fast, we use time range of $[0, 5]$. Parameters are same except for different initial and target state: Set index I to denote user 1-500 and II to denote the rest. 1st row: $\boldsymbol{x}_0 = -10, \boldsymbol{a} = 10$. 2nd row: $\boldsymbol{x}_0 = -10, \boldsymbol{a}(\mathrm{I}) = -5, \boldsymbol{a}(\mathrm{II}) = 10$. 3rd row: $\boldsymbol{x}_0$ sampled uniformly from $[-10, 10]$ and sorted in decreasing order, $\boldsymbol{a}(\mathrm{I}) = -10, \boldsymbol{a}(\mathrm{II}) = 5$. 4th row: $\boldsymbol{x}_0$ is same as (c), $\boldsymbol{a} = 10$.

# G Optimal control policy for least square opinion guiding

In this section, we derive the optimal control policy for the opinion SDE defined in (6) with the least square opinion guiding cost. First, we restate the controlled SDE in (6) as follows.

$$\mathrm{d}x_i(t) = \big(b_i + u_i(\boldsymbol{x}, t) - x_i(t)\big)\mathrm{d}t + \theta \mathrm{d}w_i(t) + \sum\nolimits_{j=1}^{U} \alpha_{ij} x_j(t) \mathrm{d}N_j(t)$$

Putting it in the vector form, we have:

$$\mathrm{d}\boldsymbol{x}(t) = \big(\boldsymbol{b} - \boldsymbol{x} + \boldsymbol{u}\big)\mathrm{d}t + \theta \mathrm{d}\boldsymbol{w}(t) + \boldsymbol{h}(\boldsymbol{x})\mathrm{d}\boldsymbol{N}(t)$$

where the $j$-th column of $\boldsymbol{h}(\boldsymbol{x})$ captures how much influence that $x_j$ has on all other users and is defined as $\boldsymbol{h}_j(\boldsymbol{x}) = \boldsymbol{B}^j \boldsymbol{x}$, where the matrix $\boldsymbol{B}^j \in \Re^{U \times U}$ and has the $j$-th column to be $(\alpha_{1j}, \cdots, \alpha_{Uj})^\top$ and zero elsewhere.

We substitute $f = \boldsymbol{b} - \boldsymbol{x}(t) + \boldsymbol{u}(,t)$, $\boldsymbol{g} = \theta$ and $\boldsymbol{h}$ to (10) and obtain the HJB equation as

$$-\frac{\partial V}{\partial t} = \min_{\boldsymbol{u}} \Big\{ \mathcal{L}(\boldsymbol{x}, \boldsymbol{u}, t) + \frac{\theta^2}{2} \mathrm{tr}\big(V_{\boldsymbol{xx}}(\boldsymbol{x}, t)\big) + V_{\boldsymbol{x}}(\boldsymbol{x}, t)^\top (\boldsymbol{b} - \boldsymbol{x}(t) + \boldsymbol{u}(t)) + \sum_{j=1}^{U} \lambda_j(t) \big( V(\boldsymbol{x} + \boldsymbol{h}_j(\boldsymbol{x}), t) - V(\boldsymbol{x}, t) \big) \Big\} \tag{42}$$

For the least square guiding problem, the instantaneous cost and terminal cost are defined as

$$\mathcal{L}(\boldsymbol{x}, \boldsymbol{u}, t) = \frac{1}{2}\|\boldsymbol{x} - \boldsymbol{a}\|^2 + \frac{1}{2}\rho\|\boldsymbol{u}\|^2, \qquad \phi(T) = \frac{1}{2}\|\boldsymbol{x}(T) - \boldsymbol{a}\|^2$$

Hence we assume that value function $V$ is quadratic in $\boldsymbol{x}$ with unknown coefficients $\boldsymbol{v}_1(t) \in \Re^U$, $\boldsymbol{v}_{11}(t) \in \Re^{U \times U}$ and $v_0(t) \in \Re$:

$$V(\boldsymbol{x}, t) = v_0(t) + \boldsymbol{v}_1(t)^\top \boldsymbol{x} + \frac{1}{2}\boldsymbol{x}^\top \boldsymbol{v}_{11}(t) \boldsymbol{x} \tag{43}$$

To find the optimal control, we substitute (43) to HJB equation and take the gradient of the right-hand side of the HJB equation (42) with respect to $\boldsymbol{u}$ and set it to $\boldsymbol{0}$. This yields the optimal feedback control policy:

$$\boldsymbol{u}^*(\boldsymbol{x}, t) = -\frac{1}{\rho} V_{\boldsymbol{x}} = -\frac{1}{\rho}\Big(\boldsymbol{v}_1(t) + \boldsymbol{v}_{11}(t)\boldsymbol{x}\Big) \tag{44}$$

Substitute $\boldsymbol{u}^*$ in (44) to the HJB equation, we first compute the four terms on the right side of the HJB equation. Note that the minimization is reached when $\boldsymbol{u} = \boldsymbol{u}^*$.

In the following derivations, we will use the property that $\boldsymbol{v}_{11} = \boldsymbol{v}_{11}^\top$ and $\boldsymbol{a}^\top \boldsymbol{b} = \boldsymbol{b}^\top \boldsymbol{a}$ for any vector $\boldsymbol{a}$ and $\boldsymbol{b}$.

The first term is:

$$\begin{aligned}
\mathcal{L}(\boldsymbol{x}, \boldsymbol{u}^*, t) &= \frac{1}{2}\boldsymbol{x}^\top \boldsymbol{x} - \boldsymbol{x}^\top \boldsymbol{a} + \frac{1}{2}\rho \boldsymbol{u}^{*\top} \boldsymbol{u}^* \\
&= \frac{1}{2}\boldsymbol{x}^\top \boldsymbol{x} - \boldsymbol{x}^\top \boldsymbol{a} + \frac{1}{2\rho}(\boldsymbol{v}_1 + \boldsymbol{v}_{11}\boldsymbol{x})^\top (\boldsymbol{v}_1 + \boldsymbol{v}_{11}\boldsymbol{x}) \\
&= \frac{1}{2}\boldsymbol{x}^\top \boldsymbol{x} - \boldsymbol{x}^\top \boldsymbol{a} + \frac{1}{2\rho}\boldsymbol{v}_1^\top \boldsymbol{v}_1 + \frac{1}{\rho}\boldsymbol{v}_1^\top \boldsymbol{v}_{11}\boldsymbol{x} + \frac{1}{2\rho}\boldsymbol{x}^\top \boldsymbol{v}_{11}\boldsymbol{v}_{11}\boldsymbol{x} \\
&= \underbrace{\frac{1}{2\rho}\boldsymbol{v}_1^\top \boldsymbol{v}_1}_{\text{scalar}} + \underbrace{\boldsymbol{x}^\top (\frac{1}{\rho}\boldsymbol{v}_{11}\boldsymbol{v}_1 - \boldsymbol{a})}_{\text{linear}} + \underbrace{\frac{1}{2}\boldsymbol{x}^\top (\frac{1}{\rho}\boldsymbol{v}_{11}\boldsymbol{v}_{11} + \boldsymbol{I})\boldsymbol{x}}_{\text{quadratic}}
\end{aligned}$$

Note that in line 1 of the expansion of $\mathcal{L}$, we dropped the constant term $\frac{1}{2}\boldsymbol{a}^\top \boldsymbol{a}$.

The second term is a scalar: $\mathrm{tr}\big(V_{\boldsymbol{xx}}(\boldsymbol{x},t) = \frac{\theta^2}{2}\mathrm{tr}(\boldsymbol{v}_{11})$. The third term is

$$
\begin{aligned}
&V_{\boldsymbol{x}}^\top(\boldsymbol{b}-\boldsymbol{x}+\boldsymbol{u}^*)\\
&=(\boldsymbol{v}_1+\boldsymbol{v}_{11}\boldsymbol{x})^\top(\boldsymbol{b}-\boldsymbol{x}-\boldsymbol{u}^*)=(\boldsymbol{v}_1+\boldsymbol{v}_{11}\boldsymbol{x})^\top(\boldsymbol{b}-\boldsymbol{x}-\frac{1}{\rho}(\boldsymbol{v}_1+\boldsymbol{v}_{11}\boldsymbol{x}))\\
&=(\boldsymbol{v}_1^\top\boldsymbol{b}-\frac{1}{\rho}\boldsymbol{v}_1^\top\boldsymbol{v}_1)-(\boldsymbol{v}_1^\top\boldsymbol{x}+\frac{1}{\rho}\boldsymbol{v}_1^\top\boldsymbol{v}_{11}\boldsymbol{x}+\frac{1}{\rho}\boldsymbol{v}_1^\top\boldsymbol{v}_{11}\boldsymbol{x}-\boldsymbol{b}^\top\boldsymbol{v}_{11}\boldsymbol{x})-\boldsymbol{x}^\top\boldsymbol{v}_{11}^\top\boldsymbol{x}-\frac{1}{\rho}\boldsymbol{x}^\top\boldsymbol{v}_{11}^\top\boldsymbol{v}_{11}\boldsymbol{x}\\
&=\underbrace{(\boldsymbol{v}_1^\top\boldsymbol{b}-\frac{1}{\rho}\boldsymbol{v}_1^\top\boldsymbol{v}_1)}_{\text{scalar}}-\underbrace{\boldsymbol{x}^\top(\boldsymbol{v}_1+\frac{2}{\rho}\boldsymbol{v}_{11}\boldsymbol{v}_1-\boldsymbol{v}_{11}\boldsymbol{b})}_{\text{linear}}-\underbrace{\frac{1}{2}\boldsymbol{x}^\top(2\boldsymbol{v}_{11}+\frac{2}{\rho}\boldsymbol{v}_{11}\boldsymbol{v}_{11})\boldsymbol{x}}_{\text{quadratic}}
\end{aligned}
$$

The fourth term is

$$
\begin{aligned}
&\sum_{j=1}^U\boldsymbol{\lambda}_j(t)(V(\boldsymbol{x}+\boldsymbol{h}_j(\boldsymbol{x}),t)-V(\boldsymbol{x},t))\\
&=\sum_{j=1}^U\boldsymbol{\lambda}_j(t)(\boldsymbol{v}_1^\top\boldsymbol{B}^j\boldsymbol{x}+\frac{1}{2}\boldsymbol{x}^\top\boldsymbol{B}^{j\top}\boldsymbol{v}_{11}\boldsymbol{B}^j\boldsymbol{x}+\frac{1}{2}\boldsymbol{x}^\top2\boldsymbol{v}_{11}\boldsymbol{B}^j\boldsymbol{x})\\
&=\underbrace{\boldsymbol{x}^\top\boldsymbol{\Lambda}^\top\boldsymbol{v}_1}_{\text{linear}}+\underbrace{\frac{1}{2}\boldsymbol{x}^\top(\sum_{j=1}^U\lambda_j\boldsymbol{B}^{j\top}\boldsymbol{v}_{11}\boldsymbol{B}^j+2\boldsymbol{v}_{11}\boldsymbol{\Lambda})\boldsymbol{x}}_{\text{quadratic}}
\end{aligned}
$$

where $\boldsymbol{\Lambda}(t)=\sum_{j=1}^U\lambda_j(t)\boldsymbol{B}^j$. Next, we compute the left side of HJB equation as:

$$-V_t=-v_0'(t)-\boldsymbol{x}^\top\boldsymbol{v}_1'(t)-\frac{1}{2}\boldsymbol{x}^\top\boldsymbol{v}_{11}'(t)\boldsymbol{x}$$

By comparing the coefficients for the scalar, linear and quadratic terms in both left-hand-side and right-hand-side of the HJB equation, we obtain three ODEs as follows.

First, only consider all the coefficients quadratic in $\boldsymbol{x}$:

$$-\boldsymbol{v}_{11}'(t)=\boldsymbol{I}+2\boldsymbol{v}_{11}(t)(-1+\boldsymbol{\Lambda}(t))+\sum_{j=1}^U\lambda_j(t)\boldsymbol{B}^{j\top}\boldsymbol{v}_{11}(t)\boldsymbol{B}^j-\frac{1}{\rho}\boldsymbol{v}_{11}(t)\boldsymbol{v}_{11}(t)$$

Second, consider the linear term:

$$-\boldsymbol{v}_1'(t)=-\boldsymbol{a}+(-1+\boldsymbol{\Lambda}^\top(t)-\frac{1}{\rho}\boldsymbol{v}_{11}(t))\boldsymbol{v}_1(t)+\boldsymbol{v}_{11}(t)\boldsymbol{b}$$

Third, consider the scalar term:

$$-v_0'(t)=\boldsymbol{b}^\top\boldsymbol{v}_1(t)+\frac{\theta^2}{2}\mathrm{tr}(\boldsymbol{v}_{11}(t))-\frac{1}{2\rho}\boldsymbol{v}_1^\top(t)\boldsymbol{v}_1(t)$$

Finally, we compute the terminal condition for the three ODEs by $V(\boldsymbol{x}(T),T)=\phi(\boldsymbol{x}(T),T)$:

$$
\begin{aligned}
V(X(T),T)&=v_0(T)+\boldsymbol{x}(T)^\top\boldsymbol{v}_1(T)+\frac{1}{2}\boldsymbol{x}(T)^\top\boldsymbol{v}_{11}(T)\boldsymbol{x}(T)\\
\phi(\boldsymbol{x}(T),T)&=-\boldsymbol{x}(T)^\top\boldsymbol{a}+\frac{1}{2}\boldsymbol{x}(T)^\top\boldsymbol{x}(T)
\end{aligned}
$$

Hence $v_0(T)=0$, $\boldsymbol{v}_1(T)=-\boldsymbol{a}$ and $\boldsymbol{v}_{11}=\boldsymbol{I}$. Note here we drop the constant term $\frac{1}{2}\boldsymbol{a}^\top\boldsymbol{a}$ in terminal cost $\phi$.

Finally, we just need to use Algorithm 1 to solve these ODEs to obtain $\boldsymbol{v}_{11}(t)$ and $\boldsymbol{v}_1(t)$. Substituting $\boldsymbol{v}_{11},\boldsymbol{v}_1$ to (44) leads to the optimal control policy.

# H Optimal control policy for opinion influence maximization

In this section, we solve the opinion influence maximization problem. The solving scheme is similar to the least square opinion shaping cost, but the derivation is different due to different cost functions.

First, we choose $\omega = 1$ and restate the controlled opinion SDE in (6) as

$$\mathrm{d}x_i(t) = \big(b_i + u_i(\boldsymbol{x}, t) - x_i(t)\big)\mathrm{d}t + \theta \mathrm{d}w_i(t) + \sum\nolimits_{j=1}^{U} \alpha_{ij} x_j(t) \mathrm{d}N_j(t)$$

Putting it in the vector form, we have:

$$\mathrm{d}\boldsymbol{x}(t) = \big(\boldsymbol{b} - \boldsymbol{x} + \boldsymbol{u}\big)\mathrm{d}t + \theta \mathrm{d}\boldsymbol{w}(t) + \boldsymbol{h}(\boldsymbol{x})\mathrm{d}\boldsymbol{N}(t)$$

where the $j$-th column of $\boldsymbol{h}(\boldsymbol{x})$ captures how much influence that $x_j$ has on all other users and is defined as $\boldsymbol{h}_j(\boldsymbol{x}) = \boldsymbol{B}^j \boldsymbol{x}$, where the matrix $\boldsymbol{B}^j \in \Re^{U \times U}$ and has the $j$-th column to be $(\alpha_{1j}, \cdots, \alpha_{Uj})^\top$ and zero elsewhere. We substitute $f = \boldsymbol{b} - \boldsymbol{x}$, $\boldsymbol{g} = \theta$ and $\boldsymbol{h}$ to (10) and obtain the HJB equation as follows

$$-\frac{\partial V}{\partial t} = \min_{\boldsymbol{u}} \Big\{ \mathcal{L}(\boldsymbol{x}, \boldsymbol{u}, t) + \frac{\theta^2}{2} \mathrm{tr}\big(V_{\boldsymbol{x}\boldsymbol{x}}(\boldsymbol{x}, t)\big) + V_{\boldsymbol{x}}(\boldsymbol{x}, t)^\top (\boldsymbol{b} - \boldsymbol{x}(t) + \boldsymbol{u}(t)) + \sum_{j=1}^{U} \lambda_j(t) \big(V(\boldsymbol{x} + \boldsymbol{h}_j(\boldsymbol{x}), t) - V(\boldsymbol{x}, t)\big) \Big\} \tag{45}$$

For opinion influence maximization, we define the cost as follows. Suppose the goal is to maximize the opinion influence at each time on $[0, T]$, the instantaneous cost $\mathcal{L}$ is defined as:

$$\mathcal{L}(\boldsymbol{x}, \boldsymbol{u}, t) = -\sum_{j=1}^{U} x_i(t) + \frac{1}{2}\|\boldsymbol{u}(t)\|^2 = -\boldsymbol{x}(t)^\top \mathbf{1} + \frac{1}{2}\|\boldsymbol{u}(t)\|^2$$

where $\mathbf{1}$ is the column vector with each entry to be one. For the terminal cost, we have: $\phi(T) = -\boldsymbol{x}(T)^\top \mathbf{1}$.

Following the similar reasoning as the least square opinion guiding problem. Since the terminal cost $\phi$ is linear in the state $\boldsymbol{x}$, the value function must be linear in $\boldsymbol{x}$, since it is the optimal value of a linear function. Hence we set the value function $V(\boldsymbol{x}, t)$ to be a linear function in $\boldsymbol{x}$ with *unknown coefficients* $\boldsymbol{v}_1(t) \in \Re^U$ and $v_0(t) \in \Re$:

$$V(\boldsymbol{x}, t) = v_0(t) + \boldsymbol{v}_1(t)^\top \boldsymbol{x} \tag{46}$$

To find the optimal control, we substitute (46) to (45) and take the gradient of the right-hand-side of (45) with respect to $\boldsymbol{u}$ and set it to $\mathbf{0}$. This yields the optimal control policy:

$$\boldsymbol{u}^*(t) = -\frac{1}{\rho} \boldsymbol{V}_{\boldsymbol{x}} = -\frac{1}{\rho} \boldsymbol{v}_1(t) \tag{47}$$

Next, we just need to compute $\boldsymbol{v}_1(t)$ to find $\boldsymbol{u}^*$. Substitute $\boldsymbol{u}^*$ in (47) to the HJB equation, we will compute the four terms on the right side of the HJB equation and derive the ODEs by comparing the coefficients. Note that the minimization is reached when $\boldsymbol{u} = \boldsymbol{u}^*$.

First, $\mathcal{L}(\boldsymbol{x}, \boldsymbol{u}^*, t)$ is expanded as:

$$\mathcal{L}(\boldsymbol{x}, \boldsymbol{u}^*, t) = -\boldsymbol{x}^\top \mathbf{1} + \frac{1}{2}\|\boldsymbol{u}^*\|^2 = \underbrace{\frac{1}{2\rho} \boldsymbol{v}_1^\top \boldsymbol{v}_1}_{\text{scalar}} - \underbrace{\boldsymbol{x}^\top \mathbf{1}}_{\text{linear}}$$

Since $V$ is linear in $\boldsymbol{x}$, $V_{\boldsymbol{x}\boldsymbol{x}} = 0$. The third term is:

$$V_{\boldsymbol{x}}^\top (\boldsymbol{b} - \boldsymbol{x} + \boldsymbol{u}^*) = \boldsymbol{v}_1^\top (\boldsymbol{b} - \boldsymbol{x} - \frac{1}{\rho} \boldsymbol{v}_1) = \underbrace{\boldsymbol{v}^\top \boldsymbol{b} - \frac{1}{\rho} \boldsymbol{v}^\top \boldsymbol{v}_1}_{\text{scalar}} - \underbrace{\boldsymbol{x}^\top \boldsymbol{v}_1}_{\text{linear}}$$

The fourth term is:

$$\sum_{j=1}^{U} \lambda_j(t) (V(\boldsymbol{x} + \boldsymbol{h}_j(\boldsymbol{x}), t) - V(\boldsymbol{x}, t)) = \sum_{j=1}^{U} \lambda_j(t) \boldsymbol{v}_1^\top \boldsymbol{h}_j(\boldsymbol{x}) = \underbrace{\boldsymbol{x}^\top \boldsymbol{\Lambda}^\top \boldsymbol{v}_1}_{\text{linear}}$$

where $\boldsymbol{\Lambda}(t) = \sum_{j=1}^{U} \lambda_j(t) \boldsymbol{B}^j$. Next, we compute the left-hand-side of HJB equation as:

$$-V_t = -v_0'(t) - \boldsymbol{x}^\top \boldsymbol{v}_1'(t) \tag{48}$$

Then by comparing the coefficients for the scalar and linear terms in both left side and right side of the HJB equation, we obtain two ODEs.

First, only consider all the coefficients linear in $\boldsymbol{x}$:

$$\boldsymbol{v}_1'(t) = \mathbf{1} + \boldsymbol{v}_1(t) - \boldsymbol{\Lambda}^\top \boldsymbol{v}_1(t) \tag{49}$$

Second, consider the linear term:

$$v_0'(t) = -\frac{1}{2\rho}\boldsymbol{v}_1^\top \boldsymbol{v}_1 - \boldsymbol{v}_1^\top \boldsymbol{b} + \frac{1}{\rho}\boldsymbol{v}_1^\top \boldsymbol{v}_1 = -\boldsymbol{v}_1(t)^\top \boldsymbol{b} + \frac{1}{2\rho}\boldsymbol{v}_1(t)^\top \boldsymbol{v}_1(t)$$

Hence we just need to solve the ODEs (48) to obtain $\boldsymbol{v}_1$ and then compute the optimal control $\boldsymbol{u}^*(t)$ from (47).

Finally we derive the terminal conditions for the above two ordinary differential equations. First, $V(T) = \phi(T) = -\boldsymbol{x}(T)^\top \mathbf{1}$ holds from the definition of the value function. Moreover, from the function form of $V$, we have $= v_0(T) + \boldsymbol{x}^\top \boldsymbol{v}_1(T)$. Hence by comparing the coefficients, we have $v_0(T) = 0$ and $\boldsymbol{v}_1(T) = -\mathbf{1}$.

With the above terminal condition and (49), we will use Algorithm 1 to solve for $\boldsymbol{v}_1(t)$ and obtain the optimal control policy.